\documentclass[10pt]{iopart}
\usepackage{iopams}
\newtheorem{thm}{Theorem}
\newtheorem{defn}{Definition}

\usepackage{multicol}
\usepackage{graphicx}
\usepackage{verbatim}
\usepackage{blkarray}
\usepackage{chngpage}
\eqnobysec
\begin{document}
\title{Quadratic invariants for discrete clusters of weakly interacting waves}

\author{Katie L Harper$^1$, Miguel D Bustamante$^2$ and Sergey V Nazarenko$^1$}

\address{$^1$ Warwick Mathematics Institute, University of Warwick, Gibbet Hill Road, Coventry CV4 7AL, UK}
\address{$^2$ School of Mathematical Sciences, University College Dublin, Belfield, Dublin 4, Ireland}

\begin{abstract}
We consider discrete clusters of quasi-resonant triads arising from a Hamiltonian three-wave equation. A cluster consists of $N$ modes forming a total of $M$ connected triads. We investigate the problem of constructing a functionally independent set of quadratic constants of motion. We show that this problem is equivalent to an underlying basic linear problem, consisting of finding the null space of a rectangular $M \times N$ matrix $\mathbb{A}$ with entries $1,$ $-1$ and $0$. In particular, we prove that the number of independent quadratic invariants is equal to $J \equiv N - M^* \geq N - M,$ where $M^*$ is the number of linearly independent rows in $\mathbb{A}.$ Thus, the problem of finding all independent quadratic invariants is reduced to a linear algebra problem in the Hamiltonian case. We establish that the properties of the quadratic invariants (e.g., locality) are related to the topological properties of the clusters (e.g., types of linkage). To do so, we formulate an algorithm for decomposing large clusters into smaller ones and show how various invariants are related to certain parts of a cluster, including the basic structures leading to $M^{*}<M.$ We illustrate our findings by presenting examples from the Charney-Hasegawa-Mima wave model, and by showing a classification of small (up to three-triad) clusters.
\end{abstract}


\section{Introduction}\label{sec: Introduction}

Let us consider a system of interacting waves where the leading order nonlinearity is quadratic. Examples include geophysical Rossby waves \cite{1} and drift waves in plasma \cite{2} both described by the Charney-Hasegawa-Mima (CHM) equation. Nonlinear interactions can be non-resonant, in the sense that the resonant condition on the frequencies is relaxed. Non-resonant interactions have been shown to be important in realistic wave systems, particularly when the amplitudes of oscillations are finite. However, in the limit of very small amplitudes only the waves that are in exact resonance remain interacting. In this paper we will deal with resonant and non-resonant cases of interacting waves (construction of quadratic conservation laws is direct in both cases), but examples will be shown only in the resonant case for simplicity.

For quadratic nonlinearity, wave interactions take place between triplets of waves which form what is known as a triad. Let us work in $d$-dimensional Fourier space with wave vectors $\mathbf{k}\in \mathbb{R}^{d}.$ A triad is made up of three modes with wave vectors $\mathbf{k}_{1}$, $\mathbf{k}_{2}$, $\mathbf{k}_{3},$ which satisfy the following three-wave conditions:
\begin{equation}
\label{eq:3-wave}
\mathbf{k}_{1}+\mathbf{k}_{2}-\mathbf{k}_{3}=0.
\end{equation}
Each wave vector $\mathbf{k}$ has an associated frequency, given by the so-called dispersion relation $\omega = \omega(\mathbf{k}).$ By definition, the triad is called `resonant' if:
\begin{equation}
\label{Resonant conditions}
\omega(\mathbf{k}_{1})+\omega(\mathbf{k}_{2})-\omega(\mathbf{k}_{3})=0.
\end{equation}
Otherwise the triad is called non-resonant or quasi-resonant.

The wave vectors $\mathbf{k}$ can either be continuous or discrete. For waves systems in an unbounded domain, the $\mathbf{k}$'s are continuous variables. Therefore, any $\mathbf{k}$ may be a member of infinitely many resonant triads. However, in this paper we will look at wave systems in bounded domains where the wave vectors are discrete variables. For simplicity, let us consider waves in a $d$-periodic box with all sides being length $L=2\pi$ and wave vectors ${\mathbf{k}}\in \mathbb{Z}^{d}.$ As a result, any $\mathbf{k}$ may now be a member of only a few non-resonant triads, and of even fewer resonant triads. Resonant and non-resonant triads which are connected via common modes can be grouped together to form {\it clusters} of various sizes ranging from ``butterflies", where two triads are joined via one mode, to a multiple-triad cluster involving a complicated network of interconnected triads. These clusters have been studied in \cite{3,4,5,6}. By definition, a cluster is called resonant if all triads within the cluster are resonant. Otherwise the cluster is called quasi-resonant or non-resonant.\\

\noindent \textbf{Types of wave turbulence}\\
There are three different regimes of wave turbulence - kinetic, discrete and mesoscopic, and these have been classified in \cite{7} and \cite{8} by different relationships between the nonlinear frequency  broadening $\Gamma$ and the frequency spacing:
\begin{equation}
\Delta\omega=\left|\frac{\partial\omega_{\mathbf{k}}}{\partial {\mathbf{k}}}\right|\frac{2\pi}{L}\sim\frac{\omega_{\mathbf{k}}}{kL}.\nonumber
\end{equation}
When wave amplitudes are very small, the nonlinear frequency broadening is much less than the frequency spacing:
\begin{equation}
\Gamma \ll \Delta\omega.\nonumber
\end{equation}
This is {\it discrete} wave turbulence and only waves that are in exact resonance can interact and exchange energy. Very large clusters are rare and there are usually a large number of small clusters, the simplest being an isolated triad. If the energy of the system is initially concentrated in these small clusters, then an energy cascade cannot take place. An extreme version of such a situation is when there are no resonant triads at all, like in the case of the capillary surface waves \cite{9}, in which case turbulence is ``frozen''. For larger amplitudes, the nonlinear frequency broadening gets bigger and originally isolated clusters may become connected via quasi-resonances, that is, resonances with small enough frequency detuning. This will allow energy to be transferred between waves which are not exactly resonant, however, this is less efficient than energy transfer between waves which are in exact resonance.

If we gradually increase the wave system's amplitudes (via the application of an external forcing or by changing the initial conditions), the nonlinear frequency broadening will eventually become sufficiently large and reach the frequency spacing:
\begin{equation}
\Gamma \sim \Delta\omega.\nonumber
\end{equation}
In this region both types of wave turbulence - discrete and kinetic, exist and the system may oscillate in time between the two regimes giving rise to a new type of wave turbulence - {\it mesoscopic} wave turbulence.

For much larger levels of forcing the resonance broadening $\Gamma$ will always greatly exceed $\Delta\omega$, in which case the wave system will be in the {\it kinetic} regime and an energy cascade between the forcing and dissipation scales gets triggered.

The description of the mesoscopic regime is one of the most important open problems of wave turbulence. Our paper is the first work to address the direct relationship between the set of independent quadratic invariants of the full-scale system and the structure of the clusters (in terms of connectivity, geometry, etc.) where the most efficient interactions take place. \\

\begin{figure}[h!]
\centering
\setlength{\unitlength}{0.75mm}
\begin{picture}(60, 45)(10, 0)
  \put(40, 35){\vector(-1, -1){20}}
  \put(40, 35){\vector(1, -1){20}}
  \put(20, 14){- - - - - - - - - - -}
  \put(18, 10){$b_{1}$}
  \put(58, 10){$b_{2}$}
  \put(38, 38){$b_{3}$}
\end{picture}
\caption{An isolated triad.}
\label{fig: An isolated triad}
\end{figure}
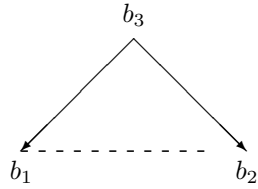

\noindent \textbf{Hamiltonian dynamical systems}\\
\newline
We consider from here on dynamical systems that are derived from a Hamiltonian equation in Fourier space:
\begin{equation}
\label{Hamiltonian}
\rmi \dot a_{\mathbf{k}}=\delta \mathcal{H}/\delta a_{\mathbf{k}}^{*},
\end{equation}
where $a_{\mathbf{k}}$ is the amplitude of the Fourier mode corresponding to the wave vector $\mathbf{k}$, $*$ denotes the complex conjugate and the Hamiltonian $\mathcal{H}$ is represented as an expansion in powers of $a_{\mathbf k}$ and $a_{\mathbf k}^{*}$:
\begin{eqnarray}
\label{Hamiltonian expansion}
\mathcal{H}&=&\mathcal{H}_{2}+\mathcal{H}_{3}+...,\\
\mathcal{H}_{2}&=&\sum\limits_{\mathbf{k}}\omega_{\mathbf{k}}|a_{\mathbf{k}}|^{2},\nonumber\\
\mathcal{H}_{3}&=&\sum\limits_{1,2,3} V_{12}^{3}a_{1}a_{2}a_{3}^{*} \, \delta_{12}^{3}+c.c.\nonumber
\end{eqnarray}
Here $a_{j}\equiv a_{\mathbf{k}_{j}}$ and $\delta_{12}^{3}\equiv \delta(\mathbf{k}_{3}-\mathbf{k}_{1}-\mathbf{k}_{2})$ is the Kronecker symbol which is one if $\mathbf{k}_{3}-\mathbf{k}_{1}-\mathbf{k}_{2}=0$  and zero otherwise. $V_{12}^{3}\equiv V({\mathbf{k}_{1},\mathbf{k}_{2}};{\mathbf{k}_{3}})$ is the nonlinear interaction coefficient.
$\mathcal{H}_{2}$ is the quadratic term and describes the non-interacting linear waves. $\mathcal{H}_{3}$ is the cubic term and describes the decaying of a single wave into two waves or the confluence of two waves into a single one. Three-wave interactions dominate wave systems with small nonlinearity provided that $\mathcal{H}_{3}\neq 0$ and the three-wave resonant conditions (\ref{Resonant conditions}) are satisfied for a non-empty set of waves. Otherwise, the leading nonlinear processes may be four-wave interactions or even higher. However, in order to fix ideas we will simply discard from here on the terms $\mathcal{H}_{4}$ and higher order, thus explicitly allowing for non-resonant interactions. Inserting $\mathcal{H}_{2}$ and $\mathcal{H}_{3}$ into (\ref{Hamiltonian expansion}) we have the evolution equation:
\begin{equation}
\label{3-wave equation}
\rmi \dot a_{\mathbf{k}}=\omega_{\mathbf{k}} a_{\mathbf{k}}+\sum\limits_{1,2}(V_{12}^{\mathbf{k}}a_{1}a_{2}\delta_{12}^{\mathbf{k}}+2V_{{\mathbf{k}}2}^{1*}\delta_{{\mathbf{k}}2}^{1}a_{2}^{*} a_{1}).
\end{equation}
In terms of an interaction representation variable, $b_{\mathbf{k}}=a_{\mathbf{k}}\rme^{\rmi\omega_{\mathbf{k}}t},$ (\ref{3-wave equation}) can be rewritten as:
\begin{equation}
\label{Int. rep. 1}
\rmi \dot b_{\mathbf{k}}=\sum\limits_{1,2} (V_{12}^{\mathbf{k}}\delta_{12}^{\mathbf{k}}b_{1}b_{2}\rme^{-\rmi \omega_{12}^{\mathbf{k}} t}+2V_{{\mathbf{k}}2}^{1*}\delta_{{\mathbf{k}}2}^{1}b_{2}^{*}b_{1}\rme^{\rmi \omega_{2 \mathbf{k}}^{1} t}),
\end{equation}
where we have introduced the so-called triad detuning parameters $\omega_{12}^{\mathbf{k}}=\omega_{{\mathbf{k}}_{1}}+\omega_{{\mathbf{k}}_{2}}-\omega_{\mathbf{k}},$ that measure the deviation of each triad from exact resonance.
This set of equations can be divided into independent subsets: the so-called quasi-resonant clusters. Within each cluster the waves interact among themselves but not with the waves of the other clusters. For example, the equations for the simplest possible cluster, consisting of one quasi-resonant triad only, are:
\begin{eqnarray}
\label{Triad}
\dot{b}_{1}&=&W^{*}b_{2}^{*}b_{3}\,\rme^{\rmi \omega_{12}^{3} t},\\
\dot{b}_{2}&=&W^{*}b_{1}^{*}b_{3}\,\rme^{\rmi \omega_{12}^{3} t},\nonumber\\
\dot{b}_{3}&=&-Wb_{1}b_{2}\,\rme^{-\rmi \omega_{12}^{3} t},\nonumber
\end{eqnarray}
where $W=2\rmi V_{12}^{3}.$ This system is integrable and its solution can be written in terms of Jacobi functions \cite{10}. This triad is represented schematically in figure \ref{fig: An isolated triad}.\\

\section{Invariants that are quadratic in the wave amplitudes}\label{sec: Linear systems of equations}

The main goal of this section is to establish that the search of all quadratic invariants of system (\ref{Int. rep. 1}) is equivalent to a much simpler problem, namely the search for the null space of a certain constant and sparse matrix, that we call ``cluster matrix''. This matrix represent the clusters of interacting triads. We show how to construct this matrix and we state and prove a theorem that establishes the equivalence with quadratic invariants.

It is important to stress that, as we will see below, finding the cluster matrix of a system of interacting triads is a straightforward matter, regardless of the size of the cluster. Moreover, finding its null space takes seconds using basic linear-algebra commands from computer programs such as Matlab. What is less trivial is that the number of these invariants is deeply related to the structure of the cluster in symbolic space, but this will be treated in the next section.\\

\noindent \textbf{Cluster Matrix and associated linear systems}\\
\newline
Let us consider a number of triads that are joined together forming a non-resonant cluster. Let the cluster consist of $M$ triads and $N$ modes, $b_n(t),\quad n = 1, \ldots, N.$ The three-wave conditions for the $m$-th triad, defined by (\ref{eq:3-wave}) can be put into the following matrix form:
\begin{equation}
{\sum\limits_{n=1}^N} A_{m n} {\mathbf{k}}_{n} = {\mathbf{0}}\,,\qquad m \,\,\mathrm{fixed}, \,\,m = 1,  \ldots , M
\end{equation}
where for each fixed $m$ the set $\{A_{m n}\}_{n=1}^N$ contains exactly two elements with value $1$, one element with value $-1,$ and the remaining elements are equal to zero. In other words, the $m$-th row of the $M \times N$ matrix $\mathbb{A} = [A_{m n}]$ corresponds to the three-wave conditions for the $m$-th triad. i.e.
\begin{equation}
A_{mn_{1}}{\mathbf{k}}_{n_{1}}+A_{mn_{2}}{\mathbf{k}}_{n_{2}}+A_{mn_{3}}{\mathbf{k}}_{n_{3}}=0,\nonumber
\end{equation}
where out of $A_{mn_{1}}, A_{mn_{2}}$ and $A_{mn_{3}}$, two have value $1$ and the third has value $-1$. From here on we refer to matrix $\mathbb{A}$ as the {\it cluster matrix}.\\

\begin{defn}
\label{def:null}
The null space of the cluster matrix $\mathbb{A}$ is the linear vector space generated from a basis of linearly independent vectors ${\bvarphi}^{(j)} \equiv (\varphi_1^{(j)} , \varphi_2^{(j)} , \ldots, \varphi_N^{(j)} )^{T}$ for which
\begin{equation}
\label{Null space}
\mathbb{A}{\bvarphi}^{(j)}={\mathbf{0}},
\end{equation}
where $j=1, \ldots, J,$ and $J$ is the dimension of the null space of $\mathbb{A}$.\\
\end{defn}

\noindent \textbf{Constructing quadratic invariants}\\
\newline
Some nonlinear problems are completely integrable - their behaviour is organized and regular; whereas non-integrable systems are not solvable exactly and exhibit chaotic behaviour. A $2n$-dimensional Hamiltonian system is said to be classically integrable in the sense of Liouville if it admits $n$ independent conserved quantities which are in involution (one can include $\mathcal{H}$ among the conserved quantities). In our case the Hamiltonian is cubic in the amplitudes, or just ``cubic'' for simplicity. If the number of independent ``quadratic'' invariants in involution is less than $n$ then we do not know whether the system is integrable or not, but we can still reduce the effective dimensionality of the system. The following theorem allows us to find all functionally independent quadratic invariants. 

\begin{thm}
\label{thm:invariants}
Consider a non-resonant cluster of $N$ interacting modes belonging to $M$ triads. Let $\varphi_n \equiv \varphi_{{\mathbf{k}}_{n}}$ be a real function of the wavenumbers of the modes in the cluster, such that the vector ${\bvarphi} \equiv (\varphi_1 , \varphi_2 , \ldots, \varphi_N)^{T}$ is in the null space of the cluster matrix: $\mathbb{A} {\bvarphi} = \mathbf{0},$ or, in components:
$\sum\limits_{n=1}^{N} A_{mn}\varphi_{n}={0}$ for all triads in the cluster, i.e., for all $m=1, \ldots, M.$ Then,
\begin{equation}
I= {\sum\limits_{n=1}^N} \varphi_n\,|b_n(t)|^2=const.,
\end{equation}
i.e. $I$ is a quadratic invariant.

Conversely, let $I = {\sum\limits_{n=1}^N} \varphi_n\,|b_n(t)|^2$ be a quadratic invariant of system (\ref{Int. rep. 1}), i.e., $\dot{I} = 0$ for all values of the complex amplitudes $b_n(t)$ such that (\ref{Int. rep. 1}) hold. Then the variables $\varphi_n$ satisfy $\sum\limits_{n=1}^{N} A_{mn}\varphi_{n}={0},$ for all $m=1, \ldots, M.$
\end{thm}

{\it Proof.}
We begin with the forward direction of proof. To show $I$ is indeed an invariant, let us take the time derivative:
\begin{equation}
\frac{\rmd I}{\rmd t}=\sum\limits_{n=1}^{N}\varphi_{n}(\dot b_{n}b_{n}^{*}+b_{n}\dot{b}_{n}^{*}).
\end{equation}
\noindent Substitute for $\dot{b_{n}}$ using (\ref{Int. rep. 1}) with $\tilde{V}_{12}^{\mathbf{k}}=V_{12}^{\mathbf{k}}\delta_{12}^{\mathbf{k}}\rme^{-\rmi \omega_{12}^{\mathbf{k}} t}$:
\begin{eqnarray}
\dot{I}&=&\sum\limits_{n=1}^{N}\varphi_{n}b_{n}^{*}\sum\limits_{1,2}^{N}(-\rmi)(\tilde{V}_{12}^{{\mathbf{k}_n}}b_{1}b_{2}+2\tilde{V}_{{\mathbf{k}}_n 2}^{1*}b_{2}^{*}b_{1})+c.c\,,\nonumber
\end{eqnarray}
where the sum over $1,2$ is a short-hand notation for a double sum over the cluster modes, and its purpose is to simplify the notation, at the expense of some abuse. Including into this double sum the sum over $n$ from 1 to $N$ in a similar way, we obtain a triple sum over modes that are exclusively in the cluster:
\begin{eqnarray}
\dot{I}&=&-\rmi\sum\limits_{1,2,3}(\varphi_{3}b_{3}^{*}b_{1}b_{2}\tilde{V}_{12}^{3}+\varphi_{1}b_{1}^{*}\tilde{V}_{12}^{3*}b_{2}^{*}b_{3}+\varphi_{2}b_{2}^{*}\tilde{V}_{12}^{3*}b_{1}^{*}b_{3}\nonumber\\
&&-\varphi_{3}b_{3}b_{1}^{*}b_{2}^{*}\tilde{V}_{12}^{3*}-\varphi_{1}b_{1}b_{2}b_{3}^{*}\tilde{V}_{12}^{3}-\varphi_{2}b_{2}b_{1}b_{3}^{*}\tilde{V}_{12}^{3})\nonumber\\
&=&-\rmi\sum\limits_{1,2,3}(\varphi_{3}-\varphi_{1}-\varphi_{2})(b_{3}^{*}b_{1}b_{2}\tilde{V}_{12}^{3}+c.c).\nonumber
\end{eqnarray}
It is clear that $\dot I=0$ because by hypothesis $\varphi_{3}-\varphi_{1}-\varphi_{2}=0$ for every term in the sum.

The converse statement follows directly from the last equation for $\dot{I}$ and the proof is omitted.\\

\noindent \textbf{Remark.} This is a general result that applies to all wave turbulence regimes (discrete, mesoscopic and kinetic). It can be seen as a generalisation of a result found in \cite{11,12}, valid for kinetic wave turbulence, on conservation laws for the three-wave kinetic equation. On the other hand, we will see in (\ref{eq:quick}) that the total number of independent invariants is equal to $J\equiv N-M^{*},$ where $M^{*}$ is the number of linearly independent triads so $M^{*} \leq M.$ It turns out that $M^{*}$ is typically greater for quasi-resonant clusters than for resonant clusters (because there are more connections in the quasi-resonant case). As a result, the number of invariants is smaller for quasi-resonant clusters. Therefore, typically in the kinetic wave turbulence and in mesoscopic wave turbulence, the only invariants that remain are the ones corresponding to the physical energy and the momenta. In some exceptional cases, there can be an extra invariant. For example, zonostrophy in Charney-Hasegawa-Mima model, as discussed in section \ref{sec: Physical invariants}.\\

\noindent \textbf{Examples.} By theorem 1, for an isolated triad described by the dynamical system (\ref{Triad}) above $I$ takes the form:
\begin{equation}
I=\varphi_{1}|b_{1}|^2+\varphi_{2}|b_{2}|^2+\varphi_{3}|b_{3}|^2.\nonumber
\end{equation}
Here, the resonant condition $\varphi_{1}+\varphi_{2}-\varphi_{3}=0$ is clearly satisfied when $\varphi_{1}=\varphi_{3}=1$, $\varphi_{2}=0$ and $\varphi_{2}=\varphi_{3}=1$, $\varphi_{1}=0$ respectively.
Therefore, there are two independent quadratic integrals of motion (called Manley-Rowe invariants):
\begin{eqnarray}
\label{Manley-Rowe}
I_{13}&=&|b_{1}|^{2}+|b_{3}|^{2},\\
I_{23}&=&|b_{2}|^{2}+|b_{3}|^{2}.\nonumber
\end{eqnarray}
A triad is integrable and the exchange of energy between the modes is periodic (phases are quasi-periodic, though). Since dynamical system (\ref{Triad}) has complex amplitudes $b_{n}(t),$ there are six variables, the real and imaginary parts of each, so for a triad to be integrable three conserved quantities are needed, namely the Hamiltonian and invariants (\ref{Manley-Rowe}) (the dynamical system can be rendered autonomous by transforming back to $a_{n}(t)$-variables).
Larger clusters may not be integrable so it is important to find a set of independent quadratic invariants. In $d$ dimensions, the most well known general quadratic invariants are the momentum components, with their density $\varphi_{n}$ equal to each of the $d$ components of the wave vector ${\mathbf{k}}.$ In the special case when the cluster is resonant, i.e., when the frequencies in each triad satisfy the resonant conditions (\ref{Resonant conditions}), we can define the energy: the quadratic invariant with density $\varphi_{n} = \omega_{n}.$  Remarkably, in the system of Rossby/drift waves one other example of $\varphi_{n}$ satisfying the resonant conditions is already known and has been discussed in the literature in the context of kinetic wave turbulence \cite{7,13,14,15}. The corresponding quadratic invariant is called {\it zonostrophy}.

\section{Properties of a cluster and its matrix $\mathbb{A}$}

There are several important relations between the cluster matrix's linear properties and the structure of the associated cluster. We enumerate them and provide some understanding:\\

\noindent \textbf{Quick counting of number of independent quadratic invariants}\\
\newline
The dimension $J (\in \mathbb{N})$ of the null space of the cluster matrix $\mathbb{A}$ (see Definition \ref{def:null}), was shown in Theorem \ref{thm:invariants} to be equal to the total number of independent quadratic invariants of the cluster system (\ref{Int. rep. 1}). By direct application of linear algebra, we have:
\begin{equation}
\label{eq:quick}
J\equiv N-M^{*}\geq N-M,
\end{equation}
where $M^{*}$ is the number of \emph{linearly independent} rows in $\mathbb{A}$.

In practice, this means that a quick counting is possible of the number of independent invariants $J$ of a given cluster. Just take the number of modes involved minus the number of triads involved, and this gives a {\it lower bound} for $J.$\\

\noindent \textbf{Connectivity of triads in a cluster and number of independent quadratic invariants}\\
\newline
There are three general results in terms of connectivity:\\

\noindent (i) In order for the triads to be connected into a cluster, the following obvious condition must be satisfied:
\begin{equation}
2M+1\geq N.\nonumber
\end{equation}
For example, consider the triple-chain in figure \ref{fig: A triple chain} below. If $2M+1<N$, then N must be greater than $7$. The only way to achieve this without adding a fourth triad to the cluster is to disconnect a triad from the chain.

\begin{figure}[h!]
\centering
\setlength{\unitlength}{0.75mm}
\begin{picture}(100, 45)(0, 0)
  \put(20, 30){\vector(-1, -1){15}}
  \put(20, 30){\vector(1, -1){15}}
  \put(50, 30){\vector(-1, -1){15}}
  \put(50, 30){\vector(1, -1){15}}
  \put(80, 30){\vector(-1, -1){15}}
  \put(80, 30){\vector(1, -1){15}}
  \put(6, 14){- - - - - - - - - - - - - - - - - - - - - - - -}
  \put(17,33){$b_{3a}$}
  \put(0, 10){$b_{1a}$}
  \put(23, 10){$b_{2a}=b_{1b}$}
  \put(47,33){$b_{3b}$}
  \put(54, 10){$b_{2b}=b_{1c}$}
  \put(77,33){$b_{3c}$}
  \put(93, 10){$b_{2c}$}
  \put(19, 20){$a$}
  \put(49, 20){$b$}
  \put(79, 20){$c$}
\end{picture}
\caption{A triple chain.}
\label{fig: A triple chain}
\end{figure}
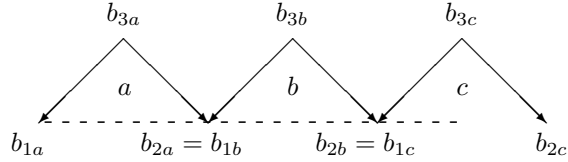

\noindent (ii) If a cluster is formed exclusively by one-common-mode connections between triads, then one has $N = 2M +1.$ Using (\ref{eq:quick}) we derive $J \geq M + 1.$
Notice that most of the exactly resonant clusters known in the literature are of this form. Therefore we expect to see the number of quadratic invariants
increase with the number of triads.\\

\noindent (iii) If a cluster is formed exclusively by two-common-mode connections between triads, then one has $N = M + 2.$ Using (\ref{eq:quick}) we derive
\begin{equation}
J \geq 2.\nonumber
\end{equation}
Notice that most of the non-resonant clusters known in the literature are of this form. In theses cases we expect to see a small number of invariants, but two of them always survive, up to the kinetic regime.

We will see below that three-common-mode connections between triads make no sense physically, so this ends our analysis in terms of connectivity properties of a cluster. In a general cluster, $N$ will be between $M+2$ and $2 M +1,$ so $J$ will vary accordingly.\\

\noindent \textbf{The dimension of the polyhedron formed by the cluster in wavenumber space}\\
\newline
Let $D$ be the dimension of the polyhedron formed by the cluster's modes in the $d$-dimensional wave vector space. In other words, $D$ is the number of linearly independent wavevectors in the set $\{\textbf{k}_n\}_{n=1}^N.$ We claim that the total number of invariants satisfies:
\begin{equation}
J \geq D.\nonumber
\end{equation}
To see this, notice first that we have $D \leq d.$ In the case $J \geq d,$ we deduce immediately $J \geq D.$ So, consider the case $J<d.$
The equations for the cluster's wave vectors are:
\begin{equation}
\label{eq:kreso}
{\sum\limits_{n=1}^N} A_{mn} \mathbf{k}_n = \mathbf{0}\,, \quad m = 1, \ldots, M,
\end{equation}
and we know there are at most $J$ independent solutions of the scalar system of equations ${\sum\limits_{n=1}^N} A_{mn} \varphi_n = 0\,, \quad m = 1, \ldots, M.$ But notice that each of the $d$ Cartesian components of (\ref{eq:kreso}) is such a scalar system of equations. Therefore, there are at most $J$ independent solutions for the Cartesian components of the $N$ wavevectors. At this point it is useful to construct a rectangular $d \times N$ matrix whose columns are the vectors $\textbf{k}_1, \textbf{k}_2, \ldots \textbf{k}_n.$ Notice that the rows of this matrix are the solutions of the $d$ Cartesian components of (\ref{eq:kreso}), so the number of linearly independent rows of this matrix is at most $J.$ On the other hand, a basic linear algebra result states that, for any matrix, the number of linearly independent rows is equal to the number of linearly independent columns. Since the number of linearly independent columns is by definition equal to $D$, the result $J \geq D$ follows directly.\\

\noindent \textbf{Examples.} Consider $d=2$ and require that resonant modes do not lie on the same line, so $D=2.$ Then $N-M^*\geq 2$, so that the solution sets of (\ref{eq:kreso}), $\{k_{x,n}\}_{n=1}^N$ and $\{k_{y,n}\}_{n=1}^N$ are allowed to be linearly independent.

Now let $d=3$ and require that resonant modes are not in the same plane, so $D=3.$ Then $N-M^*\geq 3$, so that the solution sets of (\ref{eq:kreso}), $\{k_{x,n}\}_{n=1}^N,$ $\{k_{y,n}\}_{n=1}^N$ and $\{k_{z,n}\}_{n=1}^N$ can be linearly independent.

Notice the obvious fact that for an isolated triad we have $N-M^* = 2,$ so the result is $2 \geq D$ i.e. for any host dimension $d$, the isolated triad lies either on a plane or a line.\\

\noindent \textbf{Modes belonging to only one triad.} If for a cluster $\mathbb{A}$ a mode belongs to only one triad, say triad number $m',$ then the row $A_{m'}$ corresponding to that triad is linearly independent of the other rows in $\mathbb{A}$. This is because the column corresponding to such a mode will be non-zero (1 or -1) in row $A_{m'}$ only.

\section{Physical requirements: excluded cluster matrices}\label{sec:excl}

In real-life applications and numerical simulations, we typically encounter large clusters. Sensible values of $N$ can go from $10^4$ for one-dimensional PDEs to $10^8$ for two-dimensional PDEs. For such big numbers of modes $N$ (and triads $M$), it makes sense to try to understand the basic structures appearing within a cluster, in terms of properties of the cluster matrix $\mathbb{A}.$ With this aim in mind, we present three physical requirements on the cluster matrices and their null spaces, so that the clusters represent physically sensible sets of interacting modes.

Namely, by writing out the resonant conditions for each triad from (\ref{eq:kreso}), one must admit only the matrices $A_{mn}$ for which the solution set of wavenumbers $\mathbf{k}_n\,,\quad n=1, \ldots, N,$ is physically sensible.

\begin{enumerate}
\item  The first physical requirement is that in the solution of (\ref{eq:kreso}), no two wavevectors are equal (i.e., $\mathbf{k}_{n} \neq \mathbf{k}_{n'} $ if $n \neq n'$).

    Mathematically, this requirement is summarized in the statement: We will exclude a cluster matrix $A_{mn}$ if its null space is orthogonal to any of the vectors $\mathbf{e}_i - \mathbf{e}_j,$ for some $i,j = 1, \ldots , N,$ where $\mathbf{e}_i$ is the canonical basis vector with components $(\mathbf{e}_i)_k = \delta_{i k}\,,\quad i,k = 1, \ldots, N.$

    For example, any two rows $A_m, A_{m'}$ with $m \neq m',$ must not have the same values in more than one column. The reason being that two rows having equal values in two columns would imply that the corresponding column of the third wave vector should be equal, so the two vectors would represent exactly the same triad. Therefore the following matrices are not physically sensible:\\
\begin{multicols}{2}
\begin{center}
$\left[\begin{array}{cccc}
\mathbf{1} & \mathbf{1} & -1 & 0\\
\mathbf{1} & \mathbf{1} & 0 & -1
\end{array}\right]$
\end{center}
\begin{eqnarray}
\mathbf{k}_{1}+\mathbf{k}_{2}-\mathbf{k}_{3}&=&0,\nonumber\\
\mathbf{k}_{1}+\mathbf{k}_{2}-\mathbf{k}_{4}&=&0,\nonumber
\end{eqnarray}
$$\Longrightarrow \mathbf{k}_{3}=\mathbf{k}_{4}.$$
\begin{center}
$\left[\begin{array}{cccc}
\mathbf{1} & 1 & \mathbf{-1} & 0\\
\mathbf{1} & 0 & \mathbf{-1} & 1
\end{array}\right]$
\end{center}
\begin{eqnarray}
\textbf{k}_{1}+\textbf{k}_{2}-\textbf{k}_{3}&=&0,\nonumber\\
\textbf{k}_{1}-\textbf{k}_{3}+\textbf{k}_{4}&=&0,\nonumber
\end{eqnarray}
$$\Longrightarrow \textbf{k}_{2}=\textbf{k}_{4}.$$
\end{multicols}

\item Second, if the underlying PDE is for a real function, so that there is an identification between the wavevectors $\mathbf{k}$ and $-\mathbf{k}$, then an extra requirement is that in the solution of (\ref{eq:kreso}), no two wavevectors add up to zero (i.e., $\mathbf{k}_{n} \neq - \mathbf{k}_{n'} $ if $n \neq n'$). In fact, since $b_{-{\mathbf{k}}}=b_{{\mathbf{k}}}^{*}$ one can work with only half of the $\mathbf{k}$-space for convenience, in which case possibilities to have simultaneously $\mathbf{k}$ and $-\mathbf{k}$ are automatically excluded.

Mathematically, this requirement is summarized in the statement: We will exclude a cluster matrix $A_{mn}$ if its null space is orthogonal to any of the following vectors: $\mathbf{e}_i + \mathbf{e}_j,$ for some $i,j = 1, \ldots , N.$

For example, any two rows $A_m, A_{m'}$ with $m \neq m',$ must not have values of opposite sign in two components (for example, (1,-1) for row $A_1$ and (-1,1) for row $A_2$). If this was to happen the corresponding $\mathbf{k}$  of the third non-zero component of row $A_m$ should be equal to minus the $\mathbf{k}$  of the third non-zero component of row $A_{m'},$ so one of the two vectors would be outside  the half of the $\mathbf{k}$-space, whichever way this half is selected
to describe wave fields which are real in the physical space.

Therefore the following matrix is not physically sensible:
\begin{multicols}{2}
\begin{center}
$\left[\begin{array}{cccc}
\mathbf{1} & 1 & \mathbf{-1} & 0\\
\mathbf{-1} & 0 & \mathbf{1} & 1
\end{array}\right]$
\end{center}
\begin{eqnarray}
\mathbf{k}_{1}+\mathbf{k}_{2}-\mathbf{k}_{3}&=&0,\nonumber\\
\mathbf{k}_{3}+\mathbf{k}_{4}-\mathbf{k}_{1}&=&0,\nonumber
\end{eqnarray}
$$\Longrightarrow \textbf{k}_{2}=-\textbf{k}_{4}.$$
\end{multicols}

\item Third, in cases when the zero-mode must be excluded, one must require that in the solution of (\ref{eq:kreso}), no wavevector is the zero vector (i.e., $\mathbf{k}_{n} \neq \mathbf{0} $ for all $n$).

Mathematically, this requirement is summarized in the statement: We will exclude a cluster matrix $A_{mn}$ if its null space is orthogonal to any of the following vectors: $\mathbf{e}_i,$ for some $i = 1, \ldots , N.$

We remark that in the case of triad interactions, the violation of the third requirement will imply the violation of either the first or the second requirement for some modes. To see this, notice that, for example, if
$\mathbf{k}_{1}+\mathbf{k}_{2}-\mathbf{k}_{3}=\mathbf{0}$ and $\mathbf{k}_{3}=\mathbf{0},$ then $\mathbf{k}_{1} = -\mathbf{k}_{2}.$

As an example of this third case, we have an excluded type of cluster which has the shape of a tetrahedron (see figure \ref{fig: A tetrahedron cluster}). This has the following cluster matrix:
\begin{multicols}{2}
\begin{center}
$\left[\begin{array}{cccc}
1 & -1 & 0 & 1\\
-1 & 0 & 1 & 1\\
0 & 1 & -1 & 1
\end{array}\right]$
\end{center}
\begin{eqnarray}
\mathbf{k}_{1}-\mathbf{k}_{2}+\mathbf{k}_{4}&=&0,\nonumber\\
-\mathbf{k}_{1}+\mathbf{k}_{3}+\mathbf{k}_{4}&=&0,\nonumber\\
\mathbf{k}_{2}-\mathbf{k}_{3}+\mathbf{k}_{4}&=&0,\nonumber
\end{eqnarray}
$$\Longrightarrow \mathbf{k}_{4} = 0$$
$$\Longrightarrow \mathbf{k}_{1} = \mathbf{k}_{2} = \mathbf{k}_{3}.$$
\end{multicols}

See the discussion around figure \ref{fig: A tetrahedron cluster} for more details about this unphysical case.

We remark that it is possible to construct another type of tetrahedron cluster which violates the second requirement only: this type of cluster, while not useful for the CHM model (because in CHM the underlying fields are real), could be useful in theories for complex fields. An example of this cluster is given by the following cluster matrix:
\begin{multicols}{2}
\begin{center}
$\left[\begin{array}{cccc}
1 & 1 & -1 & 0\\
0 & 1 & 1 & -1\\
-1 & 0 & 1 & 1
\end{array}\right]$
\end{center}
\begin{eqnarray}
\mathbf{k}_{1}+\mathbf{k}_{2}-\mathbf{k}_{3}&=&0,\nonumber\\
\mathbf{k}_{2}+\mathbf{k}_{3}-\mathbf{k}_{4}&=&0,\nonumber\\
-\mathbf{k}_{1}+\mathbf{k}_{3}+\mathbf{k}_{4}&=&0,\nonumber
\end{eqnarray}
$$\Longrightarrow \mathbf{k}_{4} = -\mathbf{k}_{2}.$$
\end{multicols}

\end{enumerate}

From here on, we will consider only clusters that satisfy the three physical requirements outlined above.

\section{Examples of low-dimensional clusters}\label{sec: Classification}

In this section we describe briefly clusters up to and including triple-triad clusters. 
We begin with the basic building block for all clusters - an isolated triad.

\subsection{The isolated triad}

The simplest dynamical system in the case of three wave quasi-resonances is a system of three modes, $b_{1}, b_{2}$ and $b_{3}$, called an isolated triad. It satisfies (\ref{Triad}), but we re-write it here for completeness:
\begin{figure}[h!]
\centering
\setlength{\unitlength}{0.75mm}
\begin{picture}(100, 60)(-10, 0)
  \put(40, 45){\vector(-1, -1){20}}
  \put(40, 45){\vector(1, -1){20}}
  \put(20, 24){- - - - - - - - - - -}
  \put(18, 20){$b_{1}$}
  \put(58, 20){$b_{2}$}
  \put(38, 48){$b_{3}$}
\end{picture}
\caption{An isolated triad.}
\label{fig: Another isolated triad}
\end{figure}
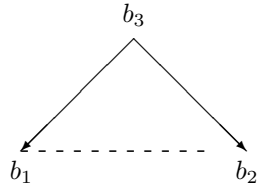\\
\begin{eqnarray}
\dot{b}_{1}&=&W^{*}b_{2}^{*}b_{3}\rme^{\Omega\,t},\\
\dot{b}_{2}&=&W^{*}b_{1}^{*}b_{3}\rme^{\Omega\,t},\nonumber\\
\dot{b}_{3}&=&-W b_{1}b_{2}\rme^{- \Omega\,t},\nonumber
\end{eqnarray}
where $W$ is the interaction coefficient and $\Omega \equiv \omega_1 + \omega_2 - \omega_3$ is the triad's detuning parameter. The cluster matrix corresponding to the resonant conditions for an isolated triad is: $$\mathbb{A}=\left[\begin{array}{ccc}
1 & 1 & -1
\end{array}\right].$$
Its null space matrix, the set of vectors ${\bvarphi}^{(j)}$ for which $\mathbb{A} {\bvarphi}^{(j)}={\mathbf{0}}$ is:
$$\Phi=\left[\begin{array}{cc}
-1 & 1\\
1 & 0\\
0 & 1
\end{array}\right].$$
A triad has $J = N - M = 2$ independent quadratic invariants. Each column of $\Phi$ gives a quadratic invariant of the dynamical system:
\begin{eqnarray}
I_{1}&=&|{b_{2}}|^{2}-|{b_{1}}|^{2},\\
I_{2}&=&|{b_{1}}|^{2}+|{b_{3}}|^{2}.\nonumber
\end{eqnarray}

Note that in a triad there are two different types of modes: for lack of a better notation, we use ``P'' for passive mode and ``A'' for active mode. For example, $b_{1}$ and $b_{2}$ are P modes and $b_{3}$ is the A mode. There is only one active mode in each triad. They correspond to substantially different scenarios of energy flux among the modes and this is discussed in \cite{3,5,6}.\\

\subsection{Double-triad clusters}

Two triads can be joined together to form a double-triad cluster. There are two main types: Butterflies (one-common-mode connection) and Kites (two-common-mode connections).\\

\noindent \textbf{Butterflies.} Two triads, $a$ and $b$, can be connected via one mode to form a butterfly. There are three different types of connection: PP, AP or AA.

A PP-butterfly consists of two triads $a$ and $b$, connected via mode $b_{1a}=b_{1b}$, which is passive in both triads.\\
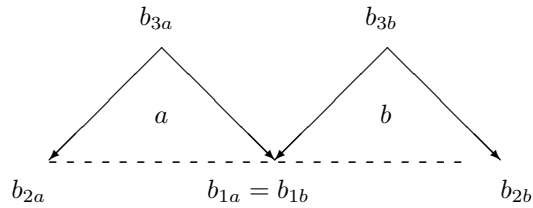
\begin{figure}[h!]
\centering
\setlength{\unitlength}{1mm}
\begin{picture}(60, 50)(5, 0)
  \put(20, 30){\vector(-1, -1){15}}
  \put(20, 30){\vector(1, -1){15}}
  \put(50, 30){\vector(-1, -1){15}}
  \put(50, 30){\vector(1, -1){15}}
  \put(5, 14){- - - - - - - - - - - - - - - - - - - - - - - -}
  \put(17,33){$b_{3a}$}
  \put(0, 10){$b_{2a}$}
  \put(26, 10){$b_{1a}=b_{1b}$}
  \put(47,33){$b_{3b}$}
  \put(65, 10){$b_{2b}$}
  \put(19, 20){$a$}
  \put(49, 20){$b$}
\end{picture}
\caption{A PP-butterfly.}
\label{fig: A PP-butterfly}
\end{figure}\\
Its dynamical system reads:
\begin{eqnarray}
\dot{b}_{1a}&=&W^{*}_{a}b_{2a}^{*}b_{3a} \rme^{\rmi \Omega_a t}+W^{*}_{b}b_{2b}^{*}b_{3b} \rme^{\rmi \Omega_b t},\\
\dot{b}_{2a}&=&W^{*}_{a}b_{1a}^{*}b_{3a} \rme^{\rmi \Omega_a t},\nonumber\\
\dot{b}_{2b}&=&W^{*}_{b}b_{1a}^{*}b_{3b} \rme^{\rmi \Omega_b t},\nonumber\\
\dot{b}_{3a}&=&-W_{a}b_{1a}b_{2a} \rme^{-\rmi \Omega_a t},\nonumber\\
\dot{b}_{3b}&=&-W_{b}b_{1a}b_{2b} \rme^{-\rmi \Omega_b t},\nonumber
\end{eqnarray}
with obvious notation for the detuning parameters of triads $a$ and $b.$ The cluster matrix is: $$\mathbb{A}=\left[\begin{array}{ccccc}
1 & 1 & -1 & 0 & 0\\
1 & 0 & 0 & 1 & -1
\end{array}\right]$$
and the null space matrix is: $$\Phi=\left[\begin{array}{ccc}
0 & -1 & 1\\
1 & 1 & -1\\
1 & 0 & 0\\
0 & 1 & 0\\
0 & 0 & 1
\end{array}\right].$$
This system has $N-M=3$ independent quadratic invariants of the form:
\begin{eqnarray}
I_{1}&=&|{b_{2a}}|^{2}+|{b_{3a}}|^{2},\label{I2}\\
I_{2}&=&|{b_{2a}}|^{2}+|{b_{2b}}|^{2}-|{b_{1a}}|^{2},\nonumber\\
I_{3}&=&|{b_{1a}}|^{2}-|{b_{2a}}|^{2}+|{b_{3b}}|^{2}.\nonumber
\end{eqnarray}
These can be linearly combined to find more quadratic invariants:
\begin{eqnarray}
I_{4}&=&|{b_{2b}}|^{2}+|{b_{3b}}|^{2},\nonumber\\
I_{5}&=&|{b_{1a}}|^{2}+|{b_{3a}}|^{2}+|{b_{3b}}|^{2}.\nonumber
\end{eqnarray}
However, only three of the five quadratic invariants above are linearly independent. A proper counting gives: 1 cubic invariant (the Hamiltonian) and 3 quadratic invariants (the ones above), totalling 4 invariants. On the other hand, the number of degrees of freedom can be reduced by noticing that 3 of the complex amplitudes' phases are slave variables, leading to 7 truly independent variables. Since there are 4 invariants for these 7 variables, at most we can reduce the system to a 3-dimensional system, which can be chaotic, as established in several papers \cite{16,17}.

Here we constructed the dynamical system for the butterfly by writing out the dynamical system for each triad a and b and substituting for the common mode, i.e. $b_{1a}=b_{1b}$. Where the two triads meet via a common mode the corresponding right hand sides are summed. This rule applies also to bigger clusters. However, from now on we will not write out the dynamical equations explicitly. Firstly, there is no need to do so since such dynamical equations can easily be reproduced from the cluster matrix $\mathbb{A}$. And secondly, for our purpose of finding invariants the cluster matrix is a more straightforward and self-sufficient approach. Likewise we will omit writing out the explicit expressions for the invariants as they can easily be produced using the null space matrix of $\mathbb{A}$. \\ 
\begin{figure}[h!]
\centering
\setlength{\unitlength}{1mm}
\begin{picture}(60, 45)(5, 0)
  \put(35, 15){\vector(-1, 0){25}}
  \put(35, 15){\vector(-1, 1){17}}
  \put(50, 30){\vector(-1, -1){15}}
  \put(50, 30){\vector(1, -1){15}}
  \put(35, 14){- - - - - - - - - - - -}
  \multiput(10, 15)(3,6){3}{\line(1,2){2}}
  \put(18,35){$b_{2a}$}
  \put(8, 10){$b_{1a}$}
  \put(26, 10){$b_{3a}=b_{1b}$}
  \put(48,35){$b_{3b}$}
  \put(65, 10){$b_{2b}$}
  \put(23, 18){$a$}
  \put(49, 20){$b$}
\end{picture}
\caption{An AP-butterfly.}
\label{fig: An AP-butterfly}
\end{figure}
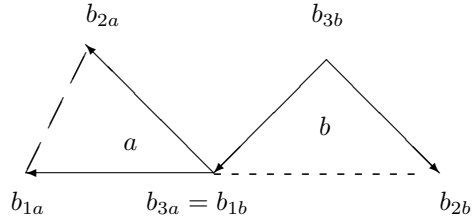
\begin{figure}[h!]
\centering
\setlength{\unitlength}{1mm}
\begin{picture}(60, 45)(5, 0)
  \put(35, 15){\vector(-1, 0){25}}
  \put(35, 15){\vector(-1, 1){17}}
  \put(35, 15){\vector(1, 0){25}}
  \put(35, 15){\vector(1, 1){17}}
  \multiput(60,15)(-3,6){3}{\line(-1,2){2}}
  \multiput(10,15)(3,6){3}{\line(1,2){2}}
  \put(18,35){$b_{2a}$}
  \put(8, 10){$b_{1a}$}
  \put(26, 10){$b_{3a}=b_{3b}$}
  \put(50,35){$b_{2b}$}
  \put(55, 10){$b_{1b}$}
  \put(23, 18){$a$}
  \put(45, 18){$b$}
\end{picture}
\caption{An AA-butterfly.}
\label{fig: An AA-butterfly}
\end{figure}
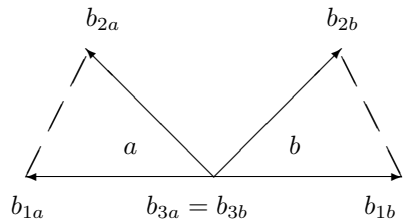\\
From now on we will omit the dashed lines from the clusters in order to make the figures less busy. The triads can easily be identified by labelling them $a,b,c$ etc., and the arrows point out off the only active mode in the triad.\\

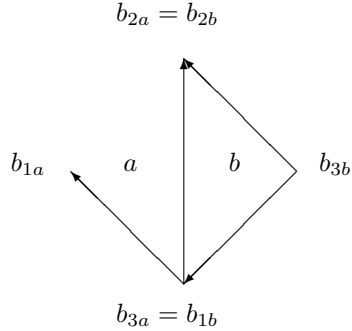
\begin{figure}[h!]
\centering
\setlength{\unitlength}{1mm}
\begin{picture}(60, 70)(0, 0)
  \put(30, 15){\vector(0, 1){30}}
  \put(30, 15){\vector(-1, 1){15}}
  \put(45, 30){\vector(-1, 1){15}}
  \put(45, 30){\vector(-1, -1){15}}
  \put(7,30){$b_{1a}$}
  \put(21, 50){$b_{2a}=b_{2b}$}
  \put(21, 10){$b_{3a}=b_{1b} $}
  \put(48, 30){$b_{3b}$}
  \put(22, 30){$a$}
  \put(36, 30){$b$}
\end{picture}
\caption{An AP-PP kite.}
\label{fig: An AP-PP kite}
\end{figure}

\noindent \textbf{Kites.} Another way of joining two triads is via two common modes, to form what is known as a kite. There is only one possible way in which to do this and that is as an AP-PP kite connected via modes $b_{3a}=b_{1b}$ and $b_{2a}=b_{2b}$ as in figure \ref{fig: An AP-PP kite}:\\
\\
The corresponding cluster matrix is: $$\mathbb{A}=\left[\begin{array}{cccc}
1 & 1 & -1 & 0\\
0 & 1 & 1 & -1
\end{array}\right].$$
It has $N-M=2$ independent quadratic invariants and the null space matrix is: $${\Phi}=\left[\begin{array}{cc}
2 & -1\\
-1 & 1\\
1 & 0\\
0 & 1
\end{array}\right].$$

To understand why certain types of kites are not realisable, we need to look at the physical requirements established in section \ref{sec:excl}. Firstly, consider a kite with the connection PP-PP; this can be shown to be wrong by considering the three-wave condition, $\mathbf{k}_{1}+\mathbf{k}_{2}=\mathbf{k}_{3}$ for both triads:\\
\begin{eqnarray}
\mathbf{k}_{1a}+\mathbf{k}_{2a}&=&\mathbf{k}_{3a},\nonumber\\
\mathbf{k}_{1b}+\mathbf{k}_{2b}&=&\mathbf{k}_{3b}.\nonumber
\end{eqnarray}
Substituting $1a=1b$ and $2a=2b$, we find that:
\begin{equation}
\mathbf{k}_{3a}=\mathbf{k}_{3b}.\nonumber
\end{equation}
This is not possible since we would be left with, $b_{1a}=b_{1b}$, $b_{2a}=b_{2b}$ and $b_{3a}=b_{3b}$ which means $a$ and $b$ are the same triad.
A similar analysis can be used to rule out the AA-PP kite, with connecting modes $b_{1a}=b_{1b}$ and $b_{3a}=b_{3b}$.

Secondly, the analysis for the kite with an AP-PA connection is more interesting. Substituting $3a=1b$ and $1a=3b$ we find that:
\begin{equation}
\mathbf{k}_{2a}=-\mathbf{k}_{2b}.\nonumber
\end{equation}
If the underlying wave field is real in the physical space, as is the case for Rossby waves and drift waves (see examples in section \ref{sec: CHM examples}) then $\mathbf{k}$ and $-\mathbf{k}$ represent the same mode via $b_{-\mathbf{k}}=b_{\mathbf{k}}^{*}$ and therefore the two triads in the kite are identical. Thus the AP-PA kite is impossible for real wave fields (but it could be possible for complex wave fields).

Finally, there cannot be kites with either connection AA-AA or AP-AP, simply because there cannot be two active modes in one triad. This has nothing to do with the physical requirements.

\subsection{Triple-triad and $N$-triad clusters}

$N$-triad clusters with $N \geq 3$ consist of $N$ triads connected via either one common mode or two common modes between any two triads. In cases of two-common-mode connections, only AP-PP connection is allowed (as seen in the previous subsection). This constrains heavily the types of connections in the clusters. For example, the physical requirements outlined in section \ref{sec:excl} imply that if two triads are joined via two common modes (necessarily with an AP-PP connection), then no other triad can share those two modes. Also, two triads cannot have three common modes.

\section{Application to resonant clusters: Charney-Hasegawa-Mima model}\label{sec: CHM examples}

Charney-Hasegawa-Mima (CHM) model describes geophysical Rossby waves and waves in magnetized plasmas. It is defined by the following equation for a real wave field $\psi$ in two dimensional physical space:
\begin{equation}
\label{CHM}
\frac{\partial}{\partial t}(\Delta\psi-F\psi)+\beta\frac{\partial\psi}{\partial x}+J[\psi,\Delta\psi]=0,
\end{equation}
where $F=1/\rho^{2}$ with $\rho$ being the Rossby deformation radius or the ion Larmor radius for Rossby and drift waves respectively and $\beta$ is the gradient of the Coriolis parameter or a measure of density gradient in plasma.

According to the CHM model (\ref{CHM}) the dispersion relation for the wave frequency is given by:
\begin{equation}
\label{Freq}
\omega_{\mathbf{k}}=-\beta k_{x}/(F+k^{2}).
\end{equation}
Since $\psi$ is a real function then $\mathbf{k}$ and $-\mathbf{k}$ represent the same mode via the property of the Fourier transform of real functions $b_{-\mathbf{k}}=b_{\mathbf{k}}^{*}$. Thus we can choose to work with only half of the Fourier space e.g. $k_{x}\geq 0$. We will further neglect $k_{x}=0$ since this corresponds to zero frequency zonal flows and these are not waves.\\

We restrict our analysis to resonant clusters, i.e., clusters such that any triad in the cluster is exactly resonant, see (\ref{Resonant conditions}). The reason for this restriction is that only resonant clusters can be written directly in the standard Hamiltonian form (\ref{Hamiltonian expansion}) with evolution equations in the standard form (\ref{Int. rep. 1}). In this particular case, $\omega_{12}^{\mathbf{k}} = 0$ for all triads so the explicit time-dependence in (\ref{Int. rep. 1}) disappears.

Let us consider separately two limiting values of the parameter $F$, corresponding physically to (1) large wavenumbers and (2) small wavenumbers.\\

\begin{figure}
\centering
\includegraphics[scale=0.7]{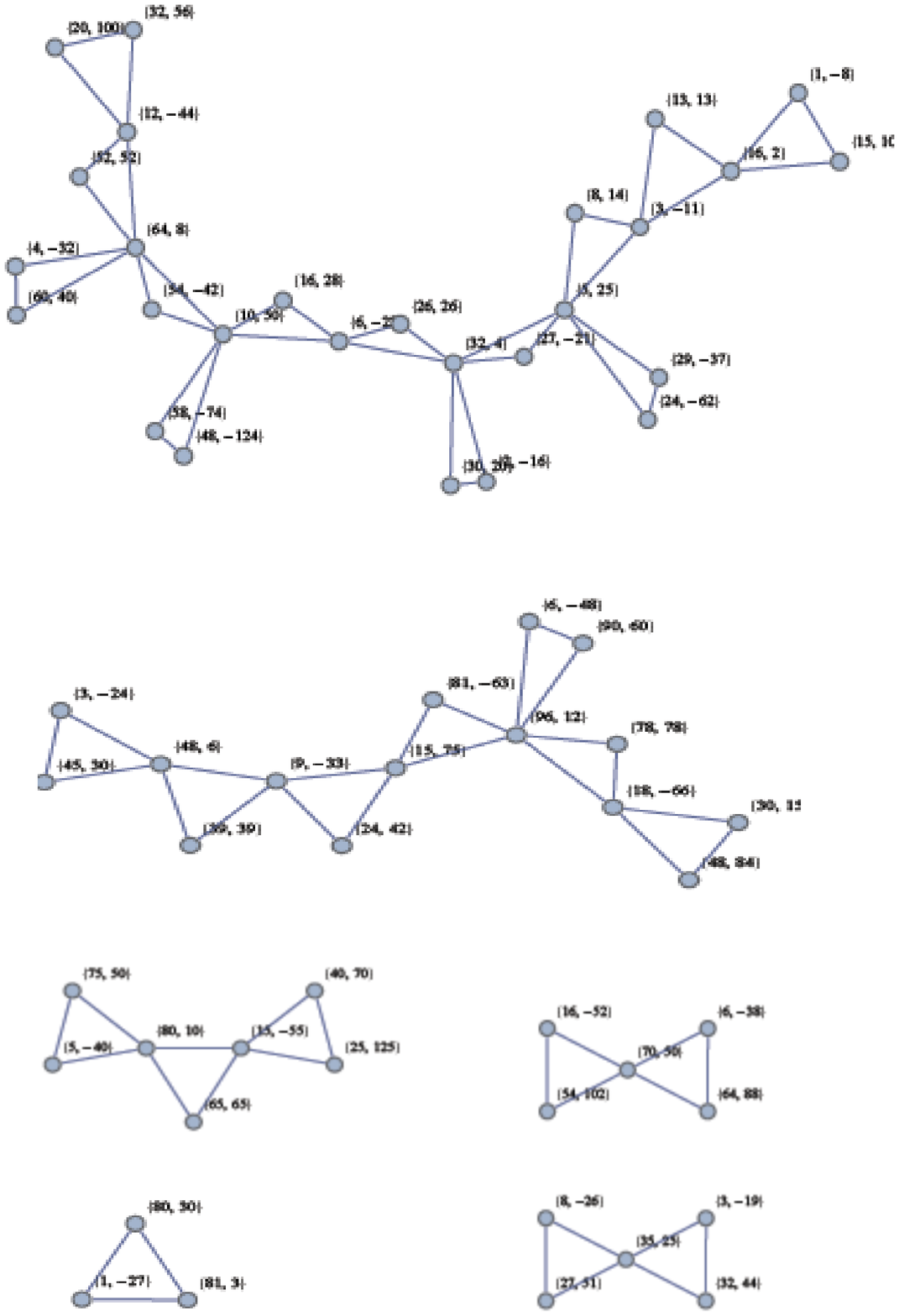}
\vspace{-1cm}
\caption{Small-scale Rossby waves in the region $1 \le k_{x}\le 100$, $-100\le k_{y}\le 100$. The mirror images have been removed and only one isolated triad is shown.}
\label{Small scale example}
\end{figure}

\noindent {\textbf{Example 1:}} Let the frequency be that of small-scale Rossby waves, $\rho k\rightarrow \infty$. The dispersion relation can be obtained by putting $F=0$ in (\ref{Freq}) which gives:
\begin{equation}
\omega_{\mathbf{k}}=-\beta k_{x}/k^{2}.
\end{equation}
If we consider the region: $1 \le k_{x}\le 100$ and $-100\le k_{y}\le 100$ we find numerically a total of thirty-four clusters (seventeen clusters plus their mirror images). This consists of twenty-four isolated triads, four butterflies, two triple-chains, two seven-triad clusters and two thirteen-triad clusters; see figure \ref{Small scale example}. It is worth noticing that this case was solved analytically very recently by one of the authors \cite{18}, and our clusters coincide with the analytical results presented in the cited reference. \\

\begin{figure}
\hspace*{-0.99in}
\includegraphics[scale=0.65]{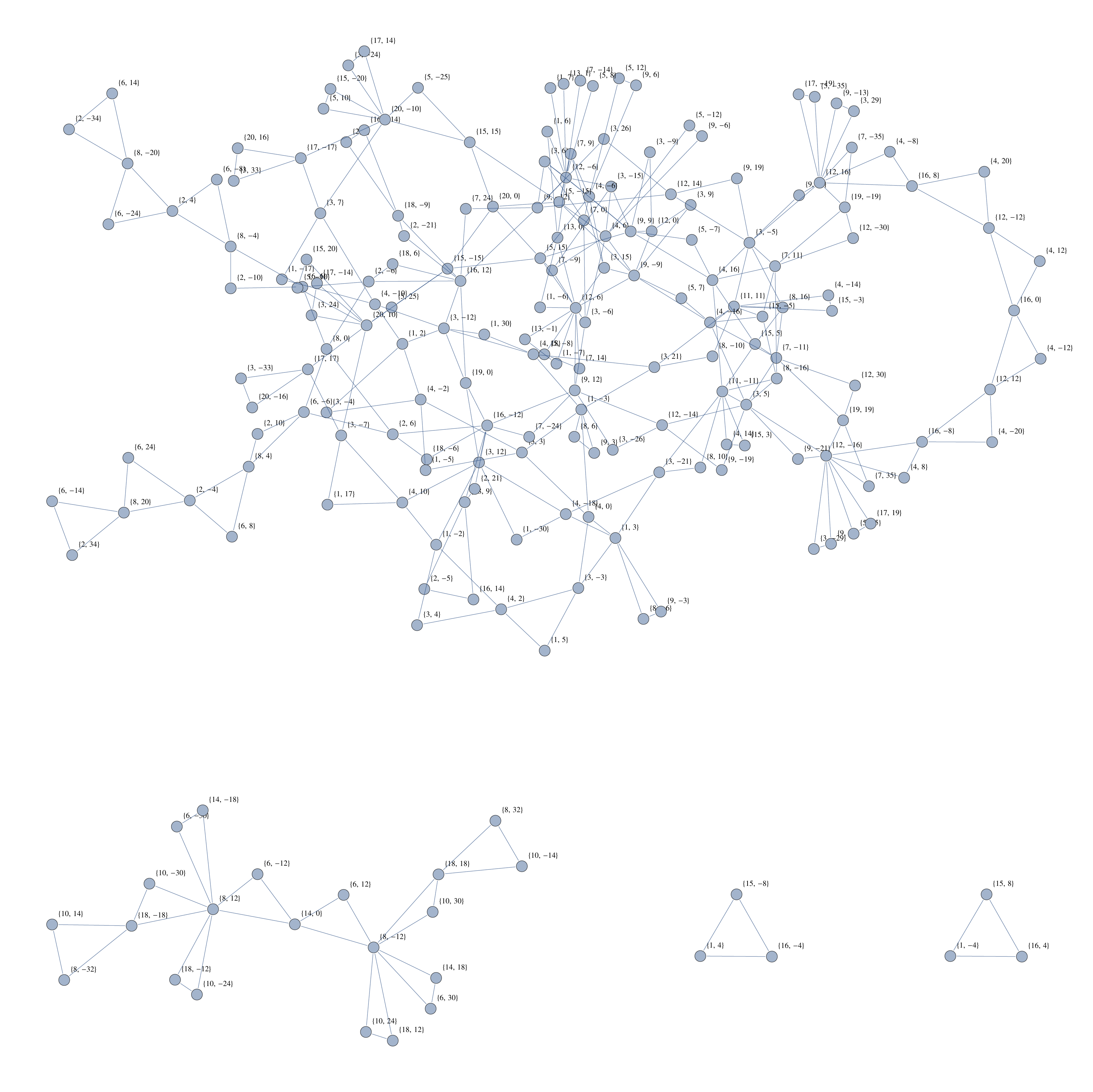}
\caption{Large-scale Rossby waves in the region $1 \le k_{x}\le 20$, $-20\le k_{y}\le 20$.}
\label{Large scale example}
\end{figure}

\noindent {\textbf{Example 2:}} Now let us consider large scale Rossby waves, $\rho^{2}k^{2}\ll 1$, with the frequency obtained from (\ref{Freq}) by Taylor expansion in $\rho^{2}k^{2}$:
\begin{equation}
\omega_{\mathbf{k}}=-\beta k_{x}\rho^{2}(1-\rho^{2}k^{2}).
\end{equation}
Since the first part in this expression is equal to ${\mathbf{k}}_{x}$, for the purpose of finding the resonances we can take a simpler expression:
\begin{equation}
\omega_{k}=k_{x}k^{2}.
\end{equation}
If we consider the region $1 \le k_{x}\le 20$ and $-20\le k_{y}\le 20$ we find numerically a total of four clusters. This consists of two isolated triads, one ten-triad cluster and one 104-triad cluster as shown in figure \ref{Large scale example}.

We see that in a much smaller domain of the large-scale limit (Example 2) we already have a much larger cluster than in the small-scale limit (Example 1). This tells us that the resonance conditions are much easier to satisfy in the large-scale limit than in the small-scale limit.


\section{Relating the quadratic invariants to the topological properties of the clusters}\label{sec: Reducing clusters}

In large clusters, not only the number of independent quadratic invariants matters. Also, the way each invariant links the different interconnected triads is important. Here we present the result that, in symbolic cluster space, typically only few invariants depend on all triads' amplitudes. The majority of the invariants are quite local in symbolic cluster space, involving one or only a few triads. Notice that being local in symbolic cluster space does not imply being local in wavenumber Fourier space. However, being non-local in symbolic cluster space implies being non-local in wavenumber Fourier space.

The algorithm presented in detail in \ref{app:reducing}, is a rather technical algorithm aimed at relating the properties of invariants (e.g., locality) to the topological properties of the clusters (e.g., types of linkage). Here we merely present the synthesis and results:\\

\noindent (i) Consider clusters for which a triad has two free modes, i.e., two of the modes belong only to that triad. As shown in part 1 of \ref{app:reducing}, such triad contributes with two quadratic invariants: one of them depends only on the two free modes (so it is ``local'') and the other invariant depends on one of the free modes as well as modes in the whole cluster. \\

\noindent (ii) Consider clusters for which a triad has only one free mode. In this case it is possible to symbolically eliminate the triad, thus ``reducing'' the original cluster into a new cluster or group of disconnected clusters, in such a way that the new cluster(s) have the same number of invariants than the original cluster. In this case the invariants of the original cluster depend explicitly on the invariants of the reduced cluster(s). Therefore, the invariants could be local or non-local, depending on the locality of the reduced cluster's invariants.\\

\noindent (iii) The above reduction procedures can be applied iteratively in order to reduce a large cluster.\\

We emphasize that our algorithm is not designed as a search procedure for the invariants: these invariants can easily be found by any symbolic algebra software (Mathematica, Matlab, etc.) via the direct computation of the null space of the cluster matrix. Instead, our algorithm is aimed at relating the properties of the invariants to the cluster topology.\\

\noindent \textbf{Example 1.} The largest cluster found in the small-scale limit of the CHM model pictured in figure \ref{Small scale example} is a $13$-triad cluster with $27$ modes. It has $J = N - M = 27 - 13 = 14$ invariants. From these, $6$ are local, each depending only on a pair of loose modes of the clusters.

The remaining $8$ invariants are relatively less local: $3$ invariants depend on $3$ modes each, and $5$ invariants depend on $4$ modes each. The linked triads can be eliminated by successively applying part 1 of our reduction algorithm.\\
In this case the cluster matrix $A$ is:\\
\newline
\begin{adjustwidth}{-1em}{}
\begin{footnotesize}
$\left[\begin{array}{ccccccccccccccccccccccccccc}
\bf{1} & \bf{1} & -1 & 0 & 0 & 0 & 0 & 0 & 0 & 0 & 0 & 0 & 0 & 0 & 0 & 0 & 0 & 0 & 0 & 0 & 0 & 0 & 0 & 0 & 0 & 0 & 0\\
0 & 0 & -1 & 1 & 1 & 0 & 0 & 0 & 0 & 0 & 0 & 0 & 0 & 0 & 0 & 0 & 0 & 0 & 0 & 0 & 0 & 0 & 0 & 0 & 0 & 0 & 0\\
0 & 0 & 0 & 0 & 1 & -1 & 1 & 0 & 0 & 0 & 0 & 0 & 0 & 0 & 0 & 0 & 0 & 0 & 0 & 0 & 0 & 0 & 0 & 0 & 0 & 0 & 0\\
0 & 0 & 0 & 0 & 0 & 0 & 1 & \bf{-1} & \bf{1} & 0 & 0 & 0 & 0 & 0 & 0 & 0 & 0 & 0 & 0 & 0 & 0 & 0 & 0 & 0 & 0 & 0 & 0\\
0 & 0 & 0 & 0 & 0 & 0 & 1 & 0 & 0 & -1 & 1 & 0 & 0 & 0 & 0 & 0 & 0 & 0 & 0 & 0 & 0 & 0 & 0 & 0 & 0 & 0 & 0\\
0 & 0 & 0 & 0 & 0 & 0 & 0 & 0 & 0 & -1 & 0 & \bf{1} & \bf{1} & 0 & 0 & 0 & 0 & 0 & 0 & 0 & 0 & 0 & 0 & 0 & 0 & 0 & 0\\
0 & 0 & 0 & 0 & 0 & 0 & 0 & 0 & 0 & -1 & 0 & 0 & 0 & 1 & 1 & 0 & 0 & 0 & 0 & 0 & 0 & 0 & 0 & 0 & 0 & 0 & 0\\
0 & 0 & 0 & 0 & 0 & 0 & 0 & 0 & 0 & 0 & 0 & 0 & 0 & 1 & 0 & -1 & 1 & 0 & 0 & 0 & 0 & 0 & 0 & 0 & 0 & 0 & 0\\
0 & 0 & 0 & 0 & 0 & 0 & 0 & 0 & 0 & 0 & 0 & 0 & 0 & 0 & 0 & 0 & 1 & \bf{-1} & \bf{1} & 0 & 0 & 0 & 0 & 0 & 0 & 0 & 0\\
0 & 0 & 0 & 0 & 0 & 0 & 0 & 0 & 0 & 0 & 0 & 0 & 0 & 0 & 0 & 0 & 1 & 0 & 0 & -1 & 1 & 0 & 0 & 0 & 0 & 0 & 0\\
0 & 0 & 0 & 0 & 0 & 0 & 0 & 0 & 0 & 0 & 0 & 0 & 0 & 0 & 0 & 0 & 0 & 0 & 0 & -1 & 0 & \bf{1} & \bf{1} & 0 & 0 & 0 & 0\\
0 & 0 & 0 & 0 & 0 & 0 & 0 & 0 & 0 & 0 & 0 & 0 & 0 & 0 & 0 & 0 & 0 & 0 & 0 & -1 & 0 & 0 & 0 & 1 & 1 & 0 & 0\\
0 & 0 & 0 & 0 & 0 & 0 & 0 & 0 & 0 & 0 & 0 & 0 & 0 & 0 & 0 & 0 & 0 & 0 & 0 & 0 & 0 & 0 & 0 & 0 & 1 & \bf{-1} & \bf{1}
\end{array}\right]$\\
\end{footnotesize}
\end{adjustwidth}
and we can verify directly that its null space matrix $\Phi$ is:
\begin{footnotesize}
$$\left[\begin{array}{cccccccccccccc}
-1 & 1 & 0 & 0 & 0 & 0 & 0 & 0 & 0 & 0 & 0 & 0& 0 & 0\\
1 & 0 & 0 & 0 & 0 & 0 & 0  & 0 & 0 & 0 & 0 & 0 & 0 & 0\\
0 & 1 & 0 & 0 & 0 & 0 & 0  & 0 & 0 & 0 & 0 & 0 & 0 & 0\\
0 & 1 & -1 & 0 & 0 & 0 & 0 & 0 & 0 & 0 & 0 & 0 & 0 & 0\\
0 & 0 & 1 & 0 & 0 & 0 & 0  & 0 & 0 & 0 & 0 & 0 & 0 & 0\\
0 & 0 & 1 & 0 & 1 & 0 & 0  & 0 & 0 & 0 & 0 & 0 & 0 & 0\\
0 & 0 & 0 & 0 & 1 & 0 & 0  & 0 & 0 & 0 & 0 & 0 & 0 & 0\\
0 & 0 & 0 & 1 & 1 & 0 & 0  & 0 & 0 & 0 & 0 & 0 & 0 & 0\\
0 & 0 & 0 & 1 & 0 & 0 & 0  & 0 & 0 & 0 & 0 & 0 & 0 & 0\\
0 & 0 & 0 & 0 & 0 & 0 & 1  & 0 & 0 & 0 & 0 & 0 & 0 & 0\\
0 & 0 & 0 & 0 & -1 & 0 & 1 & 0 & 0 & 0 & 0 & 0 & 0 & 0\\
0 & 0 & 0 & 0 & 0 & -1 & 1 & 0 & 0 & 0 & 0 & 0 & 0 & 0\\
0 & 0 & 0 & 0 & 0 & 1 & 0  & 0 & 0 & 0 & 0 & 0 & 0 & 0\\
0 & 0 & 0 & 0 & 0 & 0 & 0  & 1 & 0 & 0 & 0 & 0 & 0 & 0\\
0 & 0 & 0 & 0 & 0 & 0 & 1  & -1 & 0& 0 & 0 & 0 & 0 & 0\\
0 & 0 & 0 & 0 & 0 & 0 & 0  & 1 & 0& 1 & 0 & 0 & 0 & 0\\
0 & 0 & 0 & 0 & 0 & 0 & 0 & 0 & 0 & 1 & 0 & 0 & 0 & 0\\
0 & 0 & 0 & 0 & 0 & 0 & 0 & 0 & 1 & 1 & 0 & 0 & 0 & 0\\
0 & 0 & 0 & 0 & 0 & 0 & 0 & 0 & 1 & 0 & 0 & 0 & 0 & 0\\
0 & 0 & 0 & 0 & 0 & 0 & 0 & 0 & 0 & 0 & 0 & 1 & 0 & 0\\
0 & 0 & 0 & 0 & 0 & 0 & 0 & 0 & 0 & -1 & 0 & 1 & 0 & 0\\
0 & 0 & 0 & 0 & 0 & 0 & 0 & 0 & 0 & 0 & -1 & 1 & 0 & 0\\
0 & 0 & 0 & 0 & 0 & 0 & 0 & 0 & 0 & 0 & 1 & 0 & 0 & 0\\
0 & 0 & 0 & 0 & 0 & 0 & 0 & 0 & 0 & 0 & 0 & 1 & -1 & 0\\
0 & 0 & 0 & 0 & 0 & 0 & 0 & 0 & 0 & 0 & 0 & 0 & 1 & 0\\
0 & 0 & 0 & 0 & 0 & 0 & 0 & 0 & 0 & 0 & 0 & 0 & 1 & 1\\
0 & 0 & 0 & 0 & 0 & 0 & 0 & 0 & 0 & 0 & 0 & 0 & 0 & 1
\end{array}\right].$$\\
\end{footnotesize}

We see that all invariants in this example are relatively local within the cluster. Their existence is probably imposing severe restrictions on moving energy in and out of these triads and propagating them throughout the cluster. We could expect that invariants involving three or four modes are a bit more efficient than the invariants involving two modes, in stirring the energy through the $\mathbf k$-space. In the continuous case (kinetic wave turbulence), such efficiency happens due to the presence of the zonostrophy invariant \cite{13,15}. The presence of the zonostrophy causes anisotropy of the energy cascade which results in creation of large-scale zonal flows. Effects of the quadratic invariants onto the discrete and mesoscopic turbulence in clusters is an interesting subject for further studies.\\

\begin{figure}[h]
\centering
\includegraphics[scale=0.3]{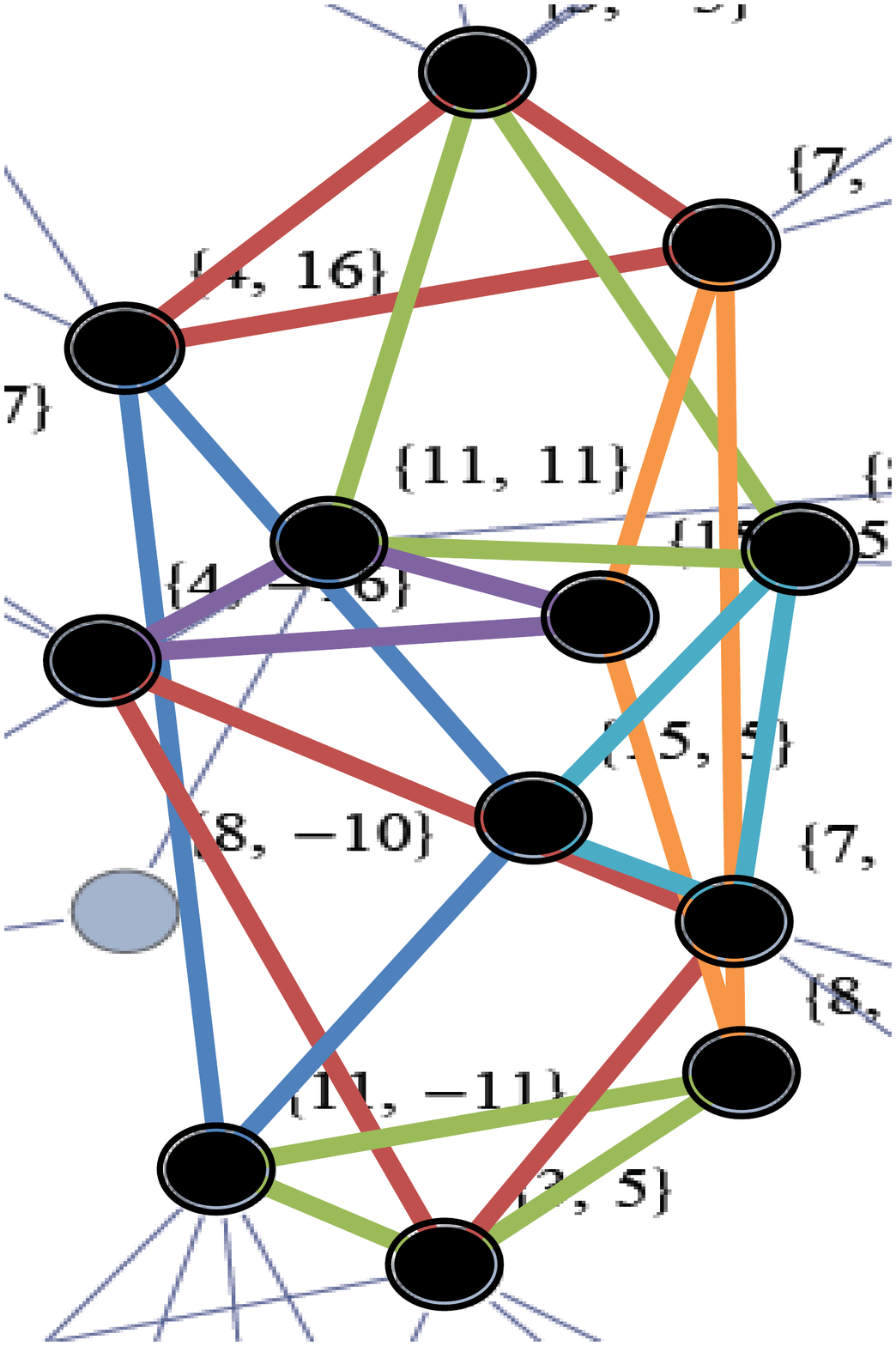}
\caption{The first cluster kernel taken from figure \ref{Large scale example} such that each triad is connected to other triads and neither part 1 or part 2 can be applied.}
\label{Cluster kernel 1}
\end{figure}

\begin{figure}[h]
\centering
\includegraphics[scale=0.3]{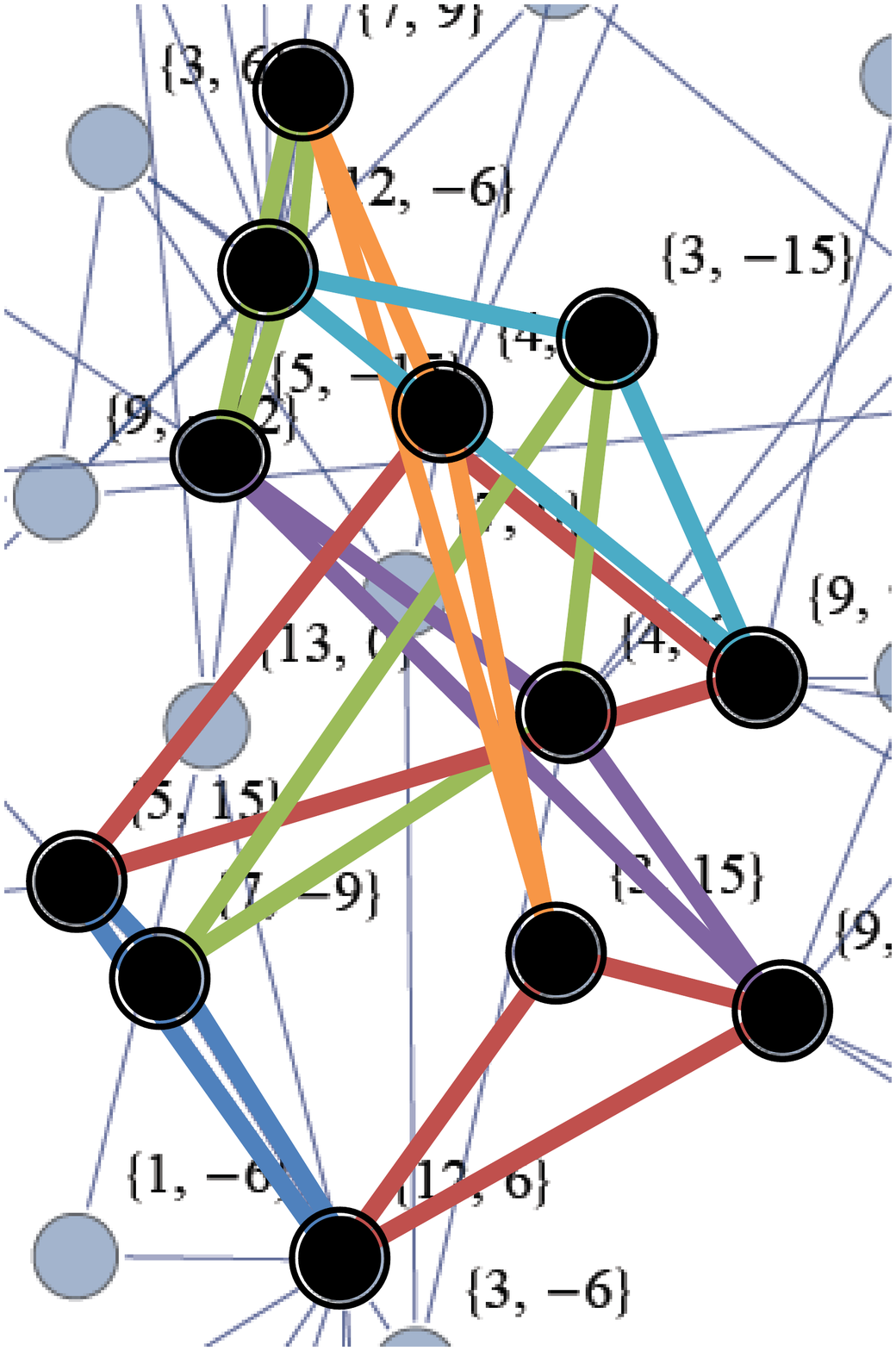}
\caption{The second cluster kernel taken from figure \ref{Large scale example}.}
\label{Cluster kernel 2}
\end{figure}

\noindent \textbf{Example 2.} Applying this procedure to the large-scale CHM example of a $104$-triad cluster ($N=178, M = 104$) in figure \ref{Large scale example}, leads to two cluster ``kernels'' after applying three times the procedure ``part 1'' in \ref{app:reducing}. The two reduced cluster kernels are made up of eight triads each and twelve modes as shown in figures \ref{Cluster kernel 1} and \ref{Cluster kernel 2}. Note that each of these clusters are mirror symmetric, i.e., each cluster maps onto itself when transformation ${\mathbf{k}}_{y}\rightarrow -{\mathbf{k}}_{y}$ is applied. The fact that both clusters are the same size is interesting but probably coincidental.
When applying our algorithm further to each of these clusters (details are given in \ref{sec-Algorithm part 3}), we conclude that for each of these ``kernel'' clusters the null space basis contains one extra vector, and so the 104-triad cluster will contain two extra invariants in total: the number of independent quadratic invariants is equal to $J=N-M^{*}= N - M + 2.$ Using $N=178, M = 104,$ we obtain $J = 178 - 104 + 2 = 76.$

\section{Physical invariants: the energy, momentum and zonostrophy}\label{sec: Physical invariants}

In this section we are going to consider the role of four physical invariants, the energy, momentum and zonostrophy, which belong to the class of quadratic invariants considered in this paper. We are also going to compare situations arising in our discrete regime (in the case of resonant clusters) to those in the kinetic regime. Firstly let us introduce briefly the kinetic regime.\\

\noindent \textbf{The kinetic regime}\\
\newline
The kinetic regime occurs when $\Gamma \gg \Delta\omega$, which is the opposite of the discrete regime, and is described by the kinetic equation (see \cite{11,12,13}):
\begin{equation}
\fl \dot{n}_{\mathbf{k}}=4\pi\int|V_{12}^{\mathbf{k}}|^{2}\delta_{12}^{\mathbf{k}}\delta(\omega_{12}^{\mathbf{k}})\times[n_{\mathbf{k}_{1}}n_{\mathbf{k}_{2}}-n_{\mathbf{k}}n_{\mathbf{k}_{1}}sign(w_{\mathbf{k}}w_{\mathbf{k}_{2}})-n_{\mathbf{k}}n_{\mathbf{k}_{2}}sign(w_{\mathbf{k}}w_{\mathbf{k}_{1}})]d\mathbf{k}_{1}d\mathbf{k}_{2},
\end{equation}
where $n_{\mathbf{k}}=\epsilon_{\mathbf{k}}/\omega_{\mathbf{k}}$ is the wave action spectrum, $V_{12}^{\textbf{k}}$ is the nonlinear interaction coefficient and $\epsilon_{\textbf{k}}$ is the energy spectrum.
It can be written in a symmetric form, since $k_{x}>0$:
\begin{equation}
\dot n_{\mathbf{k}}=\int(R_{12k}-R_{k12}-R_{2k1})d\mathbf{k}_{1}d\mathbf{k}_{2},
\end{equation}
where
\begin{equation}
R_{12k}=2\pi|V_{12}^{\mathbf{k}}|^{2}\delta_{12}^{\mathbf{k}}\delta(\omega_{12}^{\mathbf{k}})(n_{\mathbf{k}_{1}}n_{\mathbf{k}_{2}}-n_{\mathbf{k}}n_{\mathbf{k}_{1}}-n_{\mathbf{k}}n_{\mathbf{k}_{2}}).\nonumber
\end{equation}
Generally, for any quantity:
\begin{equation}
\Phi=\int \varphi_{\mathbf{k}}\dot n_{\mathbf{k}}d\mathbf{k},\nonumber
\end{equation}
with density $\varphi_{\mathbf{k}}$, it is conserved if:
\begin{equation}
\varphi_{\mathbf{k}_{3}}-\varphi_{\mathbf{k}_{1}}-\varphi_{\mathbf{k}_{2}}=0.\nonumber
\end{equation}
In other words, there is an extra invariant if the resonant relation for the density is satisfied. This is the same as for discrete wave turbulence ($\Gamma<<\Delta\omega$) even though they are very different regimes. Note that there is an intermediate regime known as mesoscopic wave turbulence in which $\Gamma \sim \Delta\omega$. This regime is much more complicated and consequently no results about additional invariants are known for this.\\

\noindent \textbf{The invariants}\\
\newline
Well-known examples of invariants are the energy and momentum with densities $\omega_{\mathbf{k}}$ and $\mathbf{k}$ respectively, and for a generic wave system no other invariant besides these has been found. However, it was discovered in \cite{13} for kinetic wave turbulence that one extra conserved quantity, independent of the energy and momentum, exists for the system of Rossby waves. This quantity is conserved under the same conditions as the kinetic equation, namely weak nonlinearity and random phases, and it cannot be conserved in interactions of higher order so may be called the invariant of the three systems. This invariant proved to be a unique additional invariant and thus the first example of wave systems with a finite number of additional invariants was obtained. This extra invariant is now known as zonostrophy.

The general expression for the density of zonostrophy, $\varsigma_{\mathbf{k}}$ was found for all ${\mathbf{k}}$'s in \cite{14}, it is:
\begin{equation}
\label{Zonostrophy general}
\varsigma_{\mathbf{k}}=\arctan\frac{k_{y}+k_{x}\sqrt{3}}{\rho k^{2}}-\arctan\frac{k_{y}-k_{x}\sqrt{3}}{\rho k^{2}}.
\end{equation}
In our paper we considered two limits- the small-scale and the large-scale limit.
In \cite{15} and \cite{19} the limit $\rho k\rightarrow \infty$ was taken in the general expression to get the density in the case of small-scale turbulence:
\begin{equation}
\label{Zonostrophy small scale}
\varsigma_{\mathbf{k}}=-\lim\limits_{\rho \rightarrow \infty}\frac{5\rho^{5}}{8\sqrt{3}}(\rho_{k}-2\sqrt{3}\omega /\beta \rho)=\frac{k_{x}^{3}}{k^{10}}(k_{x}^{2}+5k_{y}^{2}).
\end{equation}
And if we take the large-scale limit in the general expression for zonostrophy (\ref{Zonostrophy general}) above we get:
\begin{equation}
\label{Zonostrophy large scale}
\varsigma_{\mathbf{k}}=k_{x}^{3}/(k_{y}^{2}-3k_{x}^{2}).
\end{equation}

Let us consider how the energy, momentum and zonostrophy appear in our discrete clusters starting with the smallest. A triad has two linearly independent quadratic invariants (Manley-Rowe) and as a result the energy ($\omega_{\mathbf{k}}$), the two components of momentum ($k_{x}, k_{y}$) and the zonostrophy ($\varsigma_{\mathbf{k}}$) will not be linearly independent of one another. Only two may be linearly independent e.g. $k_{x}$ and $\omega_{\mathbf{k}}$ or $k_{y}$ and $\varsigma_{\mathbf{k}}$ etc.\\
To see this consider the Manley-Rowe equations:
\begin{eqnarray}
I_{1}&=&|{b_{2}}|^{2}-|{b_{1}}|^{2},\nonumber\\
I_{2}&=&|{b_{1}}|^{2}+|{b_{3}}|^{2},\nonumber
\end{eqnarray}
\begin{equation}
I=\varphi_{1}|{b_{1}}|^{2}+\varphi_{2}|{b_{2}}|^{2}+\varphi_{3}|{b_{3}}|^{2},\nonumber
\end{equation}
Substituting in $\varphi_{3}=\varphi_{1}+\varphi_{2}$ we get:
\begin{eqnarray}
I&=&\varphi_{1}(|b_{2}|^{2}+|b_{3}|^{2})+\varphi(|b_{2}|^{2}+|b_{3}|^{2}),\nonumber\\
&=&\varphi_{1}I_{1}+\varphi_{2}I_{2}.\nonumber
\end{eqnarray}

Now take a butterfly (two-triad cluster) which has three invariants in total. As a result zonostrophy does not appear as an extra invariant to the energy, and momentum components. However, any three of the four invariants will be linearly independent. Actually these considerations for a triad and a butterfly are general for all sizes. Consider larger clusters such as the triple-triad chains and stars, which both have four invariants, the zonostrophy in these cases does appear as an extra invariant as all four of $k_{x}, k_{y}, \omega_{\mathbf{k}}$ and $\varsigma_{\mathbf{k}}$ are linearly independent of one another.

Lets now consider bigger clusters arising from specific examples in the small and large scale limits. Take the biggest cluster found in the small-scale limit, shown in the top left corner of figure \ref{Small scale example}, which is made up of thirteen triads and twenty-seven modes and has fourteen linearly independent invariants. The cluster matrix $\mathbb{A}$ is shown in section  \ref{sec: Reducing clusters} from which it can be seen that triad one (row one) has two loose ends (indicated via bold print). From the null space cluster, also shown in section \ref{sec: Reducing clusters}, it is clear that triad one (column one) has a Manley-Rowe invariant. Likewise for triads 4, 6, 9, 11 and 13.

Now let us consider the energy, momentum and zonostrophy invariants in more detail in relation to the thirteen triad cluster above. Substitute the coordinates $k_{x},$ $k_{y}$ for each of the twenty-seven modes into the right hand side of (\ref{Zonostrophy small scale}) to find the values of the zonostrophy, $\varsigma_{\mathbf{k}}.$ The $x$ and $y$ momentum are simply the values of $k_{x}$ and $k_{y}$ and to get the energy values substitute $k_{x}$ and $k_{y}$ into:
\begin{equation}
\omega_{\mathbf{k}}=k_{x}/(k_{x}^{2}+k_{y}^{2}).\nonumber
\end{equation}

Firstly, check that $\omega_{\mathbf{k}}$, $k_{x}$ and $\varsigma_{\mathbf{k}}$ are in the null space of $\mathbb{A}$ and therefore are indeed invariants i.e. check that $A\omega_{\mathbf{k}}=0$, $\mathbb{A} k_{x}=0$ and $\mathbb{A} \varsigma_{\mathbf{k}}=0$. Let us now represent each of the invariants $\omega_{\mathbf{k}}$, $k_{x}$ and $\varsigma_{\mathbf{k}}$ as linear combinations of basis vectors that span the null space of $\mathbb{A}$. To do this we must find coefficient matrices $a$, $b$ and $c$ such that:
\begin{equation}
\omega=\Phi a, k_{x}=\Phi b, \varsigma=\Phi c.\nonumber
\end{equation}
Using Matlab to solve the above and rounding to three decimal places we have:
$$a=[0.046, 0.038, 0.031, 0.005, 0.023, 0.023, 0.019, 0.015, 0.003, 0.012, 0.004, 0.010, 0.008, 0.002]^{T},$$
$$b=[15, 13, 8, 24, 27, 30, 26, 16, 48, 54, 4, 52, 32, 20]^{T},$$
$$c=10^{-6} \times [0.675, 0.505, 0.450, 0.000, 0.026, 0.021, 0.016, 0.014, 0.000, 0.001,$$
$$0.000, 0.001, 0.000, 0.000]^{T}.$$
From $c$ this it is clear that the first three vectors contain the most zonostrophy and by looking at $\mathbb{A}$ and $\Phi$ in section \ref{sec: Reducing clusters} it can be seen that the first three triads in-fact contain most of the zonostrophy. This is not surprising since they have the smallest wave vectors, $\mathbf{k}$. Again, to a slightly lesser extent, it can be seen from $a$ the first three vectors also contain the most energy.

For completion let us now take an example from the large scale limit. We will consider the cluster made up of ten-triads and twenty-one modes in the bottom left corner of figure \ref{Large scale example}. The cluster matrix $\mathbb{A}$ and the null space matrix $\Phi$ are both listed in section \ref{sec: Reducing clusters}. This time to find the zonostrophy values we must substitute $k_{x}$ and $k_{y}$ into (\ref{Zonostrophy large scale}). Now find the coefficient matrix $c$ such that $\varsigma=\Phi c$:
$$c=[-9.615, 1.667, -7.043, -10.394, 6.000, 6.000, -7.043, -10.394, 1.667, 0.615, -9.615]^{T}.$$
From $c$ it is interesting to notice that the negative values in rows 1,3,4,7,8 and 11 coincide with the columns containing Manley-Rowe invariants in the null space matrix $\Phi$.

\section{Summary and Conclusions}

In this paper we consider dispersive waves involved in three-wave resonant and quasi-resonant interactions, i.e. a system of waves with quadratic nonlinearity which satisfy the 3-wave resonant conditions (\ref{eq:3-wave}) for some of the modes. We state and prove a theorem relating quadratic invariants and the wave resonance relations. It turns out that although discrete, mesoscopic and kinetic wave turbulence are different physical regimes, the conservation conditions appear to be very similar: the $\mathbf k$-space density of the quadratic invariant must satisfy the same resonance conditions as does the wave vector (and the frequency in the case of exact resonances).

In the wave turbulence of finite-sized systems, the quasi-resonant manifold splits into a set of isolated quasi-resonant clusters ranging from individual triads to much larger multiple-triad clusters. Each cluster evolves independently of the others, and therefore conservation properties hold independently for each one. In such a case of a finite-dimensional cluster the resonant condition for the density of the invariant is reduced to a linear algebra system of equations. Namely, the problem of finding invariants can be reformulated as finding the null space of the cluster matrix $\mathbb{A}$ introduced in section \ref{sec: Linear systems of equations} whose horizontal dimension $N$ is given by the number of modes and the vertical dimension $M$ is given by the number of triads.

We emphasize that the procedure of finding quadratic invariants is very easy, thanks to our formulation in terms of a linear algebra problem of finding the null space of the cluster matrix $\mathbb{A}$. So, now, the search for invariants can be done ``at the push of a button'' in any symbolic linear algebra software. Therefore, no special dedicated software (e.g., the one developed in \cite{20}) is necessary in order to search for quadratic invariants, at least in the Hamiltonian case.

We  present a classification of smaller clusters and their conservation properties up to and including three-triad clusters. We give specific examples that arise in both the small- and large-scale limits of the Charney-Hasegawa-Mima (CHM) equation. We  introduce a general algorithm for expressing quadratic invariants of large clusters in terms of quadratic invariants of smaller blocks. For example, this algorithm allows us to identify local invariants associated with individual triads independent of the rest (triads with two loose ends). The algorithm allows us to construct explicitly cases when the number of independent invariants is larger than $N-M$, explaining how these situations are related to the degeneracy of smaller blocks within the matrix (the smallest being $3\times 3$). We illustrate our algorithm by applying it to a large $104$-triad cluster arising in the large-scale CHM system, and show that it has $N-M+2$ invariants.
We also discuss, in the context of our discrete clusters, the role of well-known physical invariants, e.g. energy, momentum and zonostrophy, the extra invariant in Rossby wave systems.

Even though in this paper we only considered three-wave systems, generalisation to the four-wave and to the higher-order wave systems is straightforward. In future, it would be interesting to consider such higher-order systems, e.g. clusters of linked quartets in the system of deep water gravity waves.

Generically, quasi-resonant clusters contain more triads than resonant clusters. Thus, as the detuning in a cluster increases, $N-M$ decreases, and generally there will be less invariants. In particular, if in a cluster all triads are connected via two common modes, then the number $N-M$ is equal to $2$, corresponding to energy and enstrophy conservation (see \cite{18}). In the future, it would be interesting to study the dynamical consequences of such a loss of the quadratic invariants when triads with higher frequency detuning start to become available due to an increase of the nonlinearity of the wave turbulence.

In general, it would also be interesting to simulate numerically wave turbulence in large discrete clusters, resonant or quasi-resonant, conservative
or forced-dissipated, to see how their behaviour is different from their counterparts in kinetic wave turbulence, and how the presence of numerous
additional quadratic invariant affects the turbulent cascades.

\ack

We thank A. Cronin for enlightening discussions about linear algebra.  M.D. Bustamante acknowledges support from University College Dublin, Seed Funding Projects SF564 and SF652.
S.V. Nazarenko  acknowledges support from the government of Russian Federation  via grant
No. 12.740.11.1430 for supporting research of teams working under supervision of
invited scientists.
K.L. Harper acknowledges support from the Engineering and Physical Sciences Research Council (EPSRC).

\appendix
\section{Reducing Clusters}
\label{app:reducing}
In order to study the local character of the quadratic invariants, we need to introduce an algorithm for decomposing larger clusters into smaller clusters which completely determine the properties of the original larger clusters.\\ 
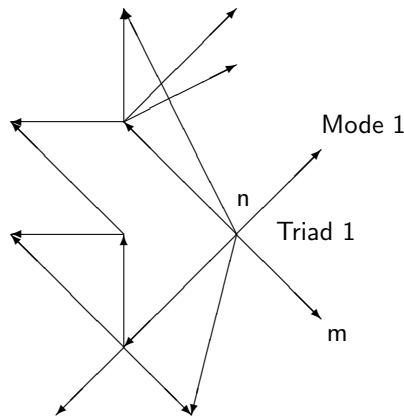
\begin{figure}[h!]
\centering
\setlength{\unitlength}{0.75mm}
\begin{picture}(90, 60)(0, 0)
  \put(45, 30){\vector(-1, 0){20}}
  \put(45, 30){\vector(-1, 1){20}}
  \put(45, 10){\vector(0, 1){20}}
  \put(45, 10){\vector(-1, 1){20}}
  \put(45, 10){\vector(1, -1){12}}
  \put(45, 10){\vector(-1, -1){12}}
  \put(45, 50){\vector(-1, 0){20}}
  \put(45, 50){\vector(0, 1){20}}
  \put(65, 30){\vector(-1, -1){20}}
  \put(65, 30){\vector(-1, 1){20}}
  \put(65, 30){\vector(-1, -4){8}}
  \put(65, 30){\vector(-1, 2){20}}
  \put(65, 30){\vector(1, 1){15}}
  \put(65, 30){\vector(1, -1){15}}
  \put(45, 50){\vector(1, 1){20}}
  \put(45, 50){\vector(2, 1){20}}
  \put(80, 48){Mode 1}
  \put(72, 29){Triad 1}
  \put(81, 11){m}
  \put(65, 35){n}
\end{picture}\\
\caption{A cluster before reduction.}
\label{fig: A cluster before reduction}
\end{figure}

\noindent\textbf{Part 1}

\begin{itemize}
\item Consider a cluster, like in figure \ref{fig: A cluster before reduction}, with an unconnected mode, let us call it mode 1 for simplicity. Being unconnected, means that mode 1 will not be joined to any other triad in the cluster other than the one it belongs to, call it triad 1.
\item Consider the cluster matrix $A$. Since mode 1 is unconnected, the rest of column 1 in the cluster matrix will contain zeros.
\item Delete column/mode one and row/triad one to leave a new reduced matrix $A'$:\\
\[A=
\begin{blockarray}{ccccccc}
1 &  &  & n &  & m & \\
\begin{block}{[ccccccc]}
1 & 0 & 0 & -1 & 0 & 1 & \cdots\\
0 & . & . & . & . & . & .\\
0 & . &  &  &  &  &\\
0 & . &  &  & A' &  &\\
0 & . &  &  &  &  &\\
0 & . &  &  &  &  &\\
\end{block}
\end{blockarray}
.\]
\item Consider a vector from the null space of $A$ (a column of matrix $\Phi$):
\[\begin{blockarray}{ccccccc}
1 &  &  & n &  & m & \\
\begin{block}{[ccccccc]}
1 & 0 & 0 & -1 & 0 & 1 & \cdots\\
0 & . & . & . & . & . & .\\
0 & . &  &  &  &  &\\
0 & . &  &  & A' &  &\\
0 & . &  &  &  &  &\\
0 & . &  &  &  &  &\\
\end{block}
\end{blockarray}
\left[ \begin{array}{c}
\varphi_{1}\\
\varphi_{2}\\
.\\
\varphi_{n}\\
.\\
\varphi_{m}\\
.\\
.\\
.\\
.\\
\varphi_{N}
\end{array} \right]=0.\]
Here $n$ and $m$ denote the positions of the non-zero entries in row one (other than the first position).
\item Solve $A'\left[\begin{array}{c}
\varphi_{2}\\
\vdots\\
\varphi_{N}
\end{array}\right]=0$, to find the null space matrix of $A'$.
\item Then solving for $\varphi_{1}$ we have:\\
\begin{equation}
\varphi_{1}-\varphi_{n}+\varphi_{m}=0 \longrightarrow
\left[ \begin{array}{c}
\varphi_{1}\\
\varphi_{2}\\
.\\
.\\
.\\
\varphi_{N}
\end{array} \right]=\left[\begin{array}{c}
\varphi_{n}-\varphi_{m}\\
\varphi_{2}\\
\vdots\\
\varphi_{N}
\end{array}\right].\label{R1}
\end{equation}
Thus finding the null space matrix of $A$ is reduced to finding the null space matrix of a smaller matrix $A'$.
By eliminating one row which is linearly independent from the rest of the rows
in $A$, and eliminating the respective column,  we have not changed the
the null space dimension. Therefore, the number of independent invariants is the same for $A$ and the smaller matrix $A'$.
\item Two situations may arise:
\begin{enumerate}
\item Triad 1 has one unconnected mode. $A'$ is a cluster matrix of a cluster obtained from $A$ by eliminating triad 1 only. It is clear that such reduced clusters will have the same number of invariants as the original bigger cluster.
\item Triad 1 has two unconnected modes e.g. $1$ and $m$ in the example below:\\
\[A=
\begin{blockarray}{ccccccc}
1 &  &  & n &  & m & \\
\begin{block}{[ccccccc]}
1 & 0 & 0 & -1 & 0 & 1 & \cdots\\
0 & . & . & . & . & 0 & .\\
0 & . &  &  &  & 0 &\\
0 & . &  & A' &  & 0 &\\
0 & . &  &  &  & 0 &\\
0 & . &  &  &  & 0 &\\
\end{block}
\end{blockarray}
.\]
In this case we have a column of zeros in matrix $A'$ (column $m$).
Thus, one can now choose $\varphi_{m}$ arbitrarily as follows,
\begin{equation}
\left[\begin{array}{c}
\varphi_{1}=\varphi_{n}-\varphi_{m}\\
\varphi_{2}\\
\vdots\\
\varphi_{m}\\
\vdots\\
\varphi_{N}
\end{array}\right]=\left[\begin{array}{c}
\varphi_{n}\\
\varphi_{2}\\
\vdots\\
0\\
\vdots\\
\varphi_{N}
\end{array}\right]+ c\left[\begin{array}{c}
-1\\
0\\
\vdots\\
1\\
\vdots\\
0
\end{array}\right].\label{R2}
\end{equation}
The second contribution here, $[-1,0,...,1,...,0]^{T}$,
gives us one of the linearly independent invariants of $A$,
and all the other columns in the null space matrix of $A$ must have zero entries in the position number $m$. Note that this  invariant is an attribute solely of the triad with two loose ends which we are eliminating, and in fact it has a simple Manley-Rowe form, $I=|b_{m}|^{2}-|b_{1}|^{2}$.

In the example above we eliminated a triad with two loose ends of type P (the ones at which the arrows point). Another possibility is when one loose end is of P type and the other is of A type (the one from which the arrows originated). It is easy to see that in this case the respective extra invariant is also of a Manley-Rowe type, $I=|b_{m}|^{2}+|b_{1}|^{2}$.\\
\end{enumerate}
\item Note that by removing triads with unconnected modes you may possibly disjoin the remaining cluster into independent clusters, which must then be treated separately.
\item Repeat until you are left with a fully connected cluster(s) i.e. one in which all modes are connected to more than one triad.
\end{itemize}

If you have completed part 1 and are left with the matrix $A'$ that cannot be reduced any further via this method move to part 2. Note however, that some clusters can be fully decomposed by repeating part 1 only and part 2 will not be necessary. This is the case for all the clusters given in the example of the small-scale Rossby waves above (figure \ref{Small scale example}). In particular the largest (thirteen-triad cluster) has six triads with two loose ends each yielding a total of six Manley-Rowe type invariants as explained above. These triads are eliminated in the first application of part 1 after which a seven-triad chain remains which will be further reduced by successive elimination of triads with double loose ends.\\

\noindent\textbf{Part 2}\\

Suppose in the remaining cluster there are two triads (triad 1 and 2 in figure \ref{fig: Cluster with no loose ends} below) that are joined together by two modes and that these modes are not connected to any other triad.\\
\begin{figure}[h!]
\centering
\setlength{\unitlength}{0.75mm}
\begin{picture}(90, 60)(0, 0)
  \put(45, 30){\vector(-1, 0){20}}
  \put(45, 30){\vector(-1, 1){20}}
  \put(45, 10){\vector(0, 1){20}}
  \put(45, 10){\vector(-1, 1){20}}
  \put(45, 50){\vector(-1, 0){20}}
  \put(45, 50){\vector(0, 1){20}}
  \put(45, 10){\vector(1, 2){10}}
  \put(45, 10){\vector(1, 0){20}}
  \put(65, 50){\vector(-1, 1){20}}
  \put(65, 50){\vector(-1, 0){20}}
  \put(55, 30){\vector(1, 2){10}}
  \put(55, 30){\vector(1, -2){10}}
  \put(35, 55){$A''$}
  \put(43, -1){$A''$}
  \put(30, 35){2}
  \put(37, 23){1}
  \put(18, 48){m}
  \put(41, 6){n}
\end{picture}
\caption{A cluster after all loose ends have been removed by part 1.}
\label{fig: Cluster with no loose ends}
\end{figure}
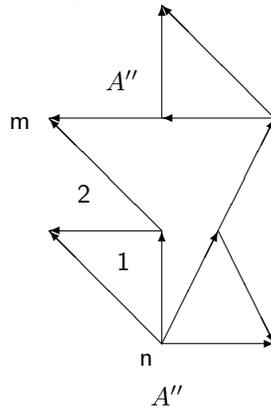\\
\begin{itemize}
\item Rearrange the rows/columns of $A'$ to form a $2\times 2$ matrix in the top left hand corner (i.e. renumber the modes in the triad in an appropriate way). The only $2\times 2$ matrix possible is $\left[\begin{array}{cc}
1 & 1\\
1 & -1
\end{array}\right]$ or any permutation of it because it must satisfy the exclusion principles meaning the following matrices are not allowed:
$\left[\begin{array}{cc}
1 & 1\\
1 & 1
\end{array}\right]$, $\left[\begin{array}{cc}
1 & -1\\
1 & -1
\end{array}\right]$, $\left[\begin{array}{cc}
1 & -1\\
-1 & 1
\end{array}\right].$ Triads 1 and 2 will form a PP-PA kite.
\item Delete the pair of modes in column one and two to get matrix $A''$ as follows:
\[A'=
\begin{blockarray}{ccccccc}
 &  &  & n & m &  & \\
\begin{block}{[ccccccc]}
1 & 1 & 0 & -1 & 0 & 0 & \cdots\\
1 & -1 & 0 & 0 & 1 & 0 & \cdots\\
0 & 0 & . & . & . & . & .\\
0 & 0 & . &  &  &  &\\
0 & 0 & . &  & A'' &  &\\
0 & 0 & . &  &  &  &\\
\end{block}
\end{blockarray}
.\]
\item By removing a pair of connected triads you may disjoin the remaining cluster into two independent clusters, in which case each must be treated separately. Or the remaining part may stay as a single cluster.
\item Solve $A''\left[\begin{array}{c}
\varphi_{3}\\
\vdots\\
\varphi_{N}
\end{array}\right]=0$, to find the null space matrix of $A''$.
\item Then solving for $\varphi_{1}$ and $\varphi_{2}$ we have:\\
$\varphi_{1}+\varphi_{2}-\varphi_{n}=0$ and $\varphi_{1}-\varphi_{2}+\varphi_{m}=0$
\begin{equation}
\longrightarrow
\left[ \begin{array}{c}
\varphi_{1}\\
\varphi_{2}\\
.\\
.\\
.\\
\varphi_{N}
\end{array} \right]=\left[\begin{array}{c}
1/2(\varphi_{n}-\varphi_{m})\\
1/2(\varphi_{n}+\varphi_{m})\\
\varphi_{3}\\
\vdots\\
\varphi_{N}
\end{array}\right].\label{R3}
\end{equation}
Therefore, the null space of $A$ is uniquely constructed from the null space of $A''$ and has the same dimension. In other words, by eliminating two triads as described above, it leads to a smaller cluster (or two disjoint clusters) whose total number of independent invariants is equal to the number of independent invariants in the original cluster.

It is not possible to have any zero columns in $A''$ since these should have been eliminated in part 1. The necessity to remove zero columns may arise only at the level of eliminating single triads and not at the level of triad pairs.
\item Look at $A''$ (single or two disjoint clusters) and again search for unconnected single modes (part 1) and triad pairs (part 2) of the type: $\left[\begin{array}{cc}
1 & 1\\
1 & -1
\end{array}\right].$ Repeat the procedure until part 1 and part 2 cannot be applied any more.\\
\end{itemize}

\noindent\textbf{Part 3}\\

After a number of successive applications of part 1 and part 2 one inevitably arrives at reduced cluster(s) for which the steps of neither part 1 nor part 2 can be carried out. Such reduced cluster(s) are usually significantly smaller than the original one but it will still have the same number of invariants. Moreover the invariants for the big cluster can be easily reconstructed from the respective invariants of such a reduced cluster by expressing the entries of the eliminated modes in the null space matrix as shown in (\ref{R1}), (\ref{R2}) and (\ref{R3}). Because of the fact that this small cluster will completely determine the conservation properties of the entire original large cluster we will call it the cluster kernel of the original cluster. Note that not all clusters will have kernels as they can be taken apart completely by the steps of part 1 and part 2.\\
\begin{itemize}
\item Let us now consider a cluster kernel (irreducible by parts 1 and 2). Following the logic of part 1 and part 2 let us now consider $3\times 3$ blocks in the top left hand corner (arising after appropriate renumbering of the rows/triads and columns/modes) such that the rest of the entries below the $3\times 3$ block contains zeros only, e.g.
$$A''=\left[\begin{array}{ccccccc}
1 & 1 & -1 & 0 & 0 & 0 & \cdots\\
1 & -1 & 0 & 0 & 1 & 0 & \cdots\\
1 & 0 & 1 & 0 & -1 & 0 & \cdots\\
0 & 0 & 0 & . & . & . & .\\
0 & 0 & 0 & . &  & A''' &\\
0 & 0 & 0 & . &  &  &
\end{array}\right].$$
Let us call these $3\times 3$ blocks $A^{3\times 3}$. Of course $A^{3\times 3}$ must again satisfy the exclusion principles discussed above. Either $A^{3\times 3}$ has:
\begin{enumerate}
\item a non-zero determinant, in which case the system of equations for $\varphi_{1}, \varphi_{2}, \varphi_{3}$ has a unique solution and consequently the complete system has the same number of invariants as $A'''$.
\item or a zero determinant, in which case, one or two independent solutions (when the $A^{3\times 3}$ rank is two or one respectively) are to be obtained by putting $\varphi_{4}=....=\varphi_{N}=0$. Further solutions are to be sought for ($\varphi_{4},....,\varphi_{N}$) given by solutions of $A^{'''}(\varphi_{4},....,\varphi_{N})^{T}=0$. For each of these solutions, the resulting system of equations for $\varphi_{1}, \varphi_{2}, \varphi_{3}$ has either:
\begin{enumerate}
\item a unique solution or
\item no solutions at all.
\end{enumerate}
By Rouch\'{e}-Capelli theorem [ref], case (2a) occurs when the rank of the coefficient matrix, $A^{3\times 3}$, in the system of equations for $\varphi_{1}, \varphi_{2}, \varphi_{3}$ is equal to the rank of the augmented matrix, $[A^{3\times 3}|\mathbf{b}]$ (where $\mathbf{b}$ depends on the fixed values of $\varphi_{4},....,\varphi_{N}$). Otherwise, if the rank of $A^{3\times 3}$ is less than the rank of $[A^{3\times 3}|\mathbf{b}]$ we will have (2b) i.e. no solutions.

Situation 2 is new with respect to $1\times 1$ and $2\times 2$ matrix eliminations above, since only starting at the $3\times 3$ matrix level can we get degenerate matrices.

In case (2a) the  system $A''$ has more invariants than $A'''$. Note that the value of $N-M$ is the same for matrix $A'''$ as for  the original matrix $A$ because the size of $A'''$ is less than the size of $A$ by the equal amount of rows and columns. This means that in case (2a) the number of linearly independent rows in the original matrix $A$, $M^{*}$ is less than the total number of rows $M$, i.e. the number of independent invariants of the full system is: $J=N-M^{*}>N-M$. An example of (2a) can be found in the following cluster:\\
\begin{figure}[h!]
\centering
\setlength{\unitlength}{0.75mm}
\begin{picture}(60, 60)(15, -5)
  \put(60, 40){\vector(-1, 0){40}}
  \put(60, 40){\vector(0, -1){25}}
  \put(60, 40){\vector(-1, -1){40}}
  \put(60, 40){\vector(1, -1){25}}
  \put(20, 0){\vector(0, 1){40}}
  \put(20, 0){\vector(1, 0){65}}
  \put(85, 15){\vector(-1, 0){25}}
  \put(85, 15){\vector(0, -1){15}}
  \put(62, 42){$b_{3a}=b_{3d}$}
  \put(11, 42){$b_{1a}=b_{1b}$}
  \put(11, -5){$b_{3b}=b_{1d}$}
  \put(78, -5){$b_{2b}=b_{1c}$}
  \put(87, 15){$b_{2a}=b_{3c}$}
  \put(51, 10){$b_{2c}=b_{2d}$}
\end{picture}
\caption{A cluster demonstrating case (2a) where $M^{*}<M$.}
\label{fig: M*<M}
\end{figure}
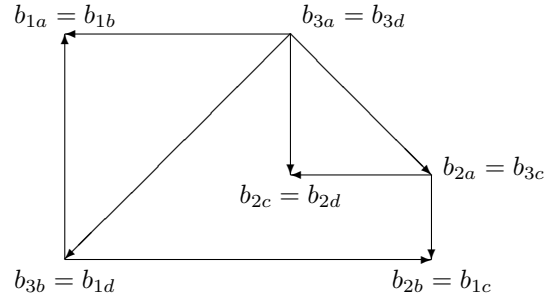\\
It has the cluster matrix: $$A''=\left[\begin{array}{cccccc}
1 & 1 & -1 & 0 & 0 & 0\\
1 & 0 & 0 & 1 & -1 & 0\\
0 & -1 & 0 & 1 & 0 & 1\\
0 & 0 & -1 & 0 & 1 & 1
\end{array}\right].$$\\
This cluster is fully connected and cannot be reduced by part 1 or part 2.
Rearrange $A''$ to form a $3\times 3$ matrix, $A^{3\times 3}$, in the top left corner:
$$A''=\left[\begin{array}{ccc|ccc}
1 & 1 & 0 & -1 & 0 & 0\\
1 & 0 & 1 & 0 & -1 & 0\\
0 & -1 & 1 & 0 & 0 & 1\\ \hline
0 & 0 & 0 & -1 & 1 & 1
\end{array}\right].$$
The determinant of $A^{3\times 3}$ is zero, so let us use the Rouch\'{e}-Capelli theorem. The rank of $A^{3\times 3}$ is two. To apply the theorem we must find the vector $\mathbf{b}$. From $A'''$ (the lower-right part in $A''$) it can be seen that $\varphi_{4}=\varphi_{5}+\varphi_{6}$. So we have two independent solutions: either $\varphi_{4}=\varphi_{5}=1,\varphi_{6}=0$ or $\varphi_{4}=\varphi_{6}=1,\varphi_{5}=0$.\\
To find $\mathbf{b}$ corresponding to these solutions, multiply the rectangular matrix in the top-right of $A^{3\times 3}$ in $A''$ by $(\varphi_{4},\varphi_{5},\varphi_{6})^{T}$. So either: $$\mathbf{b}_{1}=\left[ \begin{array}{ccc}
-1 & 0 & 0\\
0 & -1 & 0\\
0 & 0 & 1
\end{array} \right]\left[ \begin{array}{c}
1\\
1\\
0
\end{array} \right]=\left[ \begin{array}{c}
-1\\
-1\\
0
\end{array} \right],$$ or $$\mathbf{b}_{2}=\left[ \begin{array}{ccc}
-1 & 0 & 0\\
0 & -1 & 0\\
0 & 0 & 1
\end{array} \right]\left[ \begin{array}{c}
1\\
0\\
1
\end{array} \right]=\left[ \begin{array}{c}
-1\\
0\\
1
\end{array} \right].$$
Using these in the augmented matrix $[A^{3\times 3}|\mathbf{b}]$ it can be seen that the rank is again two. By the Rouch\'{e}-Capelli theorem, this system of equations has an infinite number of solutions. Hence, the number of linearly independent rows $M^{*}=3$ is less than the number of rows $M=4$ and hence $A$ has an extra invariant.\\
\newline
This explains why the null space matrix for $A$, $\Phi=\left[\begin{array}{ccc}
-1 & 1 & 0\\
1 & 0 & 1\\
1 & 0 & 0\\
0 & 1 & 1\\
0 & 1 & 0\\
0 & 0 & 1
\end{array}\right],$ doesn't satisfy the rule $J=N-M=2$ but instead $J=N-M^{*}=3$ i.e. $J=N-M^{*}\geq N-M.$

In case (2b) we ``loose'' solutions, i.e. we may find that extra solutions gained by solving for $\varphi_{1}, \varphi_{2}, \varphi_{3}$ with $\varphi_{4}=....=\varphi_{N}=0$ may be compensated by an equal or larger loss because some solutions ($\varphi_{4},....,\varphi_{N}$) of $A'''(\varphi_{4},....,\varphi_{N})^{T}=0$ do not correspond to any solution of the full system $A$. Therefore in case (2b) we may have the number of independent solutions of the original cluster $A$ to be the same or less than $A'''$. To see this take the following ``tetrahedron cluster'':\\
\begin{figure}[h!]
\centering
\setlength{\unitlength}{0.65mm}
\begin{picture}(50, 45)(-10, -5)
  \put(20, 45){\line(0, -1){25}}
  \put(-5, -5){\line(1, 1){25}}
  \put(45, -5){\line(-1, 1){25}}
  \put(-5, -5){\line(1, 0){50}}
  \put(20, 45){\line(-1, -2){25}}
  \put(45, -5){\line(-1, 2){25}}
  \put(12, 20){$a$}
  \put(20, 5){$b$}
  \put(25, 20){$c$}
\end{picture}
\caption{A tetrahedron cluster.}
\label{fig: A tetrahedron cluster}
\end{figure}
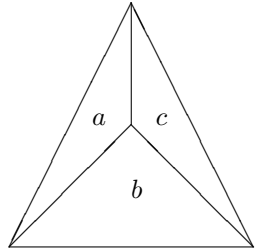\\

This cluster is unphysical (since the resulting null space is such that $\mathbf{k}_4=\mathbf{0}$, thus violating the third physical requirement in section \ref{sec:excl}) but for illustrating case (2b) it is a simple example to consider.

The cluster matrix which has been rearranged to form a $3\times 3$ matrix, $A^{3\times 3}$, on the left hand side and a column, $\mathbf{b}$, on the right is: $$A''=\left[\begin{array}{ccc|c}
1 & -1 & 0 & 1\\
-1 & 0 & 1 & 1\\
0 & 1 & -1 & 1\\
\end{array}\right].$$
The determinant of $A^{3\times 3}$ is zero. Now in order to solve our system of equations:
$$\left[\begin{array}{cccc}
1 & -1 & 0 & 1\\
-1 & 0 & 1 & 1\\
0 & 1 & -1 & 1\\
\end{array}\right]
\left[\begin{array}{c}
\varphi_{1}\\
\varphi_{2}\\
\varphi_{3}\\
\varphi_{4}
\end{array}\right]=0,$$
let us form an augmented matrix: $\left[A^{3\times 3}|\mathbf{b}\right]$ with $\mathbf{b}=(\varphi_{4},\varphi_{4},\varphi_{4})$, i.e. we have:
$$A^{3\times 3}\left[\begin{array}{c}
\varphi_{1}\\
\varphi_{2}\\
\varphi_{3}
\end{array}\right] +\varphi_{4}\mathbf{b}=0.$$
Since there is no $A'''$ in this case, we can choose $\varphi_{4}$ arbitrarily e.g. $\varphi_{4}=1$. The rank of the coefficient matrix $A^{3\times 3}$ is two and the rank of the augmented matrix $[A^{3\times 3}|\mathbf{b}]$ is three. Hence, by the Rouch\'{e}-Capelli theorem no solutions exist.\\
Now let $\varphi_{4}=0$ and solve $A^{3\times 3}(\varphi_{1}\varphi_{2},\varphi_{3})^{T}=0$. Since the rank of $A^{3\times 3}$ is two, we have one independent solution, $(\varphi_{1}\varphi_{2},\varphi_{3})=(1,1,1).$\\
Consequently, the tetrahedron cluster has $N-M=1$ invariant and it's null space matrix is:
$$\Phi=\left[\begin{array}{c}
1\\
1\\
1\\
0
\end{array}\right].$$
Therefore, for the tetrahedron cluster the number of invariants corresponds to the ``$N-M$'' rule (which holds for non-degenerate cases) even though its $A^{3\times 3}$ matrix is degenerate.
\end{enumerate}
\end{itemize}

\section{Cluster kernels of the 104-triad cluster in the large-scale CHM model}\label{sec-Algorithm part 3}

Let us consider the giant 104-triad cluster (``frog") of the large-scale CHM model shown in figure \ref{Large scale example}.
As we said in the main text, by three successive part-1 steps of our reduction algorithm this cluster can be reduced to the cluster kernels shown
in  figures \ref{Cluster kernel 1} and \ref{Cluster kernel 2}.
Both of these kernels appear to be so tightly linked that no further reduction is possible by removing triad pairs (kites), triple- or even four-triad blocks. This brings us straight to considering $5\times5$ blocks.

\noindent \textbf{Figure \ref{Cluster kernel 1}:} $A''$ has been rearranged to form a $5\times 5$ matrix in the top left hand corner:\\
$$\left[\begin{array}{ccccc|ccccccc}
1 & 1 & -1 & 0 & 0 & 0 & 0 & 0 & 0 & 0 & 0 & 0\\
1 & 0 & 0 & 0 & -1 & 0 & 0 & 0 & 1 & 0 & 0 & 0\\
0 & 1 & 0 & 0 & 0 & 1 & -1 & 0 & 0 & 0 & 0 & 0\\
0 & 0 & 0 & -1 & 1 & 0 & 0 & 1 & 0 & 0 & 0 & 0\\
0 & 0 & 1 & -1 & 0 & 0 & 0 & 0 & 0 & 1 & 0 & 0\\\hline
0 & 0 & 0 & 0 & 0 & 0 & -1 & 0 & 1 & 0 & 1 & 0\\
0 & 0 & 0 & 0 & 0 & 0 & 0 & 1 & 0 & 0 & -1 & 1\\
0 & 0 & 0 & 0 & 0 & -1 & 0 & 0 & 0 & 1 & 0 & 1
\end{array}\right]$$
The determinant of $A^{5\times 5}$ is zero and the rank is four. Now find the vector $\mathbf{b}$ from $A'''$:
\begin{eqnarray}
-\varphi_{7}+\varphi_{9}+\varphi_{11}&=&0,\\
\varphi_{8}-\varphi_{11}+\varphi_{12}&=&0,\nonumber\\
-\varphi_{6}+\varphi_{10}+\varphi_{12}&=&0.\nonumber
\end{eqnarray}
One independent solution is $\varphi_{7}=\varphi_{8}=\varphi_{9}=\varphi_{10}=1$ and $\varphi_{12}=-1$ and $\varphi_{6}=\varphi_{11}=0$. So $$\mathbf{b}=\left[\begin{array}{ccccccc}
0 & 0 & 0 & 0 & 0 & 0 & 0\\
0 & 0 & 0 & 1 & 0 & 0 & 0\\
1 & -1 & 0 & 0 & 0 & 0 & 0\\
0 & 0 & 1 & 0 & 0 & 0 & 0\\
0 & 0 & 0 & 0 & 1 & 0 & 0
\end{array}\right]\left[\begin{array}{c}
0\\
1\\
1\\
1\\
1\\
0\\
-1
\end{array}\right]=\left[\begin{array}{c}
0\\
1\\
-1\\
1\\
1
\end{array}\right].$$
The rank of $[A^{5\times 5}|\mathbf{b}]$ is four. So by the Rouch\'{e}-Capelli theorem the cluster kernel in figure \ref{Cluster kernel 1} has an infinite number of solutions and since the rank of the coefficient matrix is one less than its size, one extra invariant. The null space for figure \ref{Cluster kernel 1} is:$${\Phi}=\left[\begin{array}{ccccc}
-1 & 1 & 0 & 0 & 1\\
1 & 0 & -1 & 0 & -1\\
0 & 1 & -1 & 0 & 0\\
1 & 0 & 0 & -1 & 0\\
0 & 1 & 0 & -1 & 1\\
0 & 0 & 1 & 0 & 1\\
1 & 0 & 0 & 0 & 0\\
0 & 0 & 0 & 1 & -1\\
0 & 1 & 0 & 0 & 0\\
0 & 0 & 1 & 0 & 0\\
0 & 0 & 0 & 1 & 0\\
0 & 0 & 0 & 0 & 1
\end{array}\right].$$
\newline
\textbf{Figure \ref{Cluster kernel 2}:} Once again $A''$ has been rearranged to form a $5\times 5$ matrix in the top left hand corner:\\
$$\left[\begin{array}{ccccc|ccccccc}
1 & 1 & -1 & 0 & 0 & 0 & 0 & 0 & 0 & 0 & 0 & 0\\
-1 & 0 & 0 & 1 & 0 & 1 & 0 & 0 & 0 & 0 & 0 & 0\\
0 & 0 & -1 & 0 & 1 & 0 & 1 & 0 & 0 & 0 & 0 & 0\\
0 & 1 & 0 & 0 & 0 & 0 & 0 & 1 & -1 & 0 & 0 & 0\\
0 & 0 & 0 & 1 & -1 & 0 & 0 & 0 & 0 & 1 & 0 & 0\\\hline
0 & 0 & 0 & 0 & 0 & 0 & 0 & 0 & 0 & 1 & 1 & -1\\
0 & 0 & 0 & 0 & 0 & 0 & 1 & 1 & 0 & 0 & -1 & 0\\
0 & 0 & 0 & 0 & 0 & 1 & 0 & 0 & 1 & 0 & 0 & -1
\end{array}\right]$$
The determinant of $A^{5\times 5}$ is zero and the rank is four. Now find the vector $\mathbf{b}$ from $A'''$:
\begin{eqnarray}
\varphi_{10}+\varphi_{11}-\varphi_{12}=0,\\
\varphi_{7}+\varphi_{8}-\varphi_{11}=0,\nonumber\\
\varphi_{6}+\varphi_{9}-\varphi_{12}=0.\nonumber
\end{eqnarray}
One independent solution is $\varphi_{6}=\varphi_{7}=\varphi_{10}=\varphi_{12}=1$ and $\varphi_{8}=-1$ and $\varphi_{9}=\varphi_{11}=0$. So $$\mathbf{b}=\left[\begin{array}{ccccccc}
0 & 0 & 0 & 0 & 0 & 0 & 0\\
1 & 0 & 0 & 0 & 0 & 0 & 0\\
0 & 1 & 0 & 0 & 0 & 0 & 0\\
0 & 0 & 1 & -1 & 0 & 0 & 0\\
0 & 0 & 0 & 0 & 1 & 0 & 0
\end{array}\right]\left[\begin{array}{c}
1\\
1\\
-1\\
0\\
1\\
0\\
1
\end{array}\right]=\left[\begin{array}{c}
0\\
1\\
1\\
-1\\
1
\end{array}\right].$$
The rank of $[A^{5\times 5}|\mathbf{b}]$ is four. So by the Rouch\'{e}-Capelli theorem the cluster kernel in figure \ref{Cluster kernel 2} has an infinite number of solutions and since the rank of the coefficient matrix is one less than its size, one extra invariant. The null space for figure \ref{Cluster kernel 2} is:$${\Phi}=\left[\begin{array}{ccccc}
1 & 0 & -1 & 1 & 0\\
0 & -1 & 1 & 0 & 0\\
1 & -1 & 0 & 1 & 0\\
1 & 0 & 0 & 1 & -1\\
0 & 0 & -1 & 0 & 1\\
1 & 0 & 0 & 0 & 0\\
0 & -1 & 0 & 1 & 0\\
0 & 1 & 0 & 0 & 0\\
0 & 0 & 1 & 0 & 0\\
0 & 0 & 0 & -1 & 1\\
0 & 0 & 0 & 1 & 0\\
0 & 0 & 0 & 0 & 1
\end{array}\right].$$

Thus, both kernels have an additional invariant each. Therefore, the original 104-triad cluster has two extra invariants, $ J=N-M+2 = 178-104+2 =76$.

\end{document}